\documentclass[manuscript=article]{achemso}            		

\setkeys{acs}{usetitle=true}

\usepackage{graphicx}
\usepackage{subfigure}
\usepackage{multirow}
\usepackage{chemformula}
\let\ce\ch
\usepackage{longtable}
\usepackage{adjustbox}

\usepackage[utf8]{inputenc} 
\usepackage{booktabs}
\usepackage{multirow}
\usepackage{lscape}
\usepackage{caption}
\usepackage{xcolor}

\title{The Intercalation Chemistry of the Disordered RockSalt \ce{Li3V2O5} Anode from Cluster Expansions and Machine Learning Interatomic Potentials}
\author{Xingyu Guo}
\affiliation[University of California San Diego]{Materials Science and Engineering Program, University of California San Diego, 9500 Gilman Dr, Mail Code 0448, La Jolla, CA 92093-0448, United States}
\author{Chi Chen}
\affiliation[University of California San Diego]
{Department of Nanoengineering, University of California San Diego, La Jolla, CA, USA}
\author{Shyue Ping Ong}
\affiliation[University of California San Diego]
{Department of Nanoengineering, University of California San Diego, La Jolla, CA, USA}
\email{ongsp@eng.ucsd.edu}

\begin{document}
\maketitle

\begin{abstract}
Disordered rocksalt (DRX) \ce{Li3V2O5} is a promising candidate for anode in rechargeable lithium-ion batteries because of its ideal low voltage, high rate capability, and superior cycling stability. Herein, we presents a comprehensive study of intercalation chemistry of the DRX-\ce{Li3V2O5} anode using density functional theory calculations combined with machine learning cluster expansions and interatomic potentials. The predicted voltage profile of the disordered \ce{Li3V2O5} anode at room temperature based on Monte Carlo simulations with a fitted cluster expansion model is in excellent agreement with experiments. In contrast to previous DFT results, we find that Li ions predominately intercalate into tetrahedral sites during charging, while the majority of Li and V ions at octahedral sites remain stable. In addition, MD simulations with a fitted moment tensor potential attribute the fast-charging capability of DRX-\ce{Li3V2O5} to the facile diffusivity of \ce{Li+} via tetrahedral - octahedral - tetrahedral pathway. We further suggest tuning the Li:V ratio as a means to trade off increased lithiation capacity and decreased anode voltage in this system. This work provides in-depth insights into the high-performance DRX-\ce{Li3V2O5} anode, and paves the way to the discovery of other disordered anode materials.

\end{abstract}
		
\section{Introduction}

Rocksalt oxides have been extensively studied as electrodes for rechargeable lithium-ion batteries (LIBs).\cite{clement2020cation} As the name implies, the \ce{O2-} anions in rocksalt oxides are arranged in a close-packed face-centered-cubic (fcc) framework, with the cations occupying the tetrahedral and octahedral interstitial sites, as shown in Figure\ref{fig:pd_0K}(a). For instance, the common layered transition metal (M) oxide \ce{LiMO2} cathode used in LIBs is formed by an ordered arrangement of Li and M in this framework. In the past decade or so, lithium-rich disordered rocksalt (DRX) oxides have emerged as a promising class of alternative electrode materials with extraordinarily high specific capacities and high rate capabilities.\cite{lee2014unlocking,urban2016computational,clement2020cation, kitchaev2018design}

While DRX materials have been extensively studied as cathodes, relatively few have been explored as anodes.\cite{liu2020disordered,barnes2022electrochemically,xiong2011amorphous} Among the most promising is the DRX-\ce{Li3V2O5} recently reported by \citet{liu2020disordered}. The DRX-\ce{Li3V2O5} anode operates at a near-optimal average voltage of $\sim0.6$ V vs \ce{Li/Li^+} - high enough to alleviate the safety concerns attributed to Li plating that occurs during fast charge/discharge of the commercial graphite anode used in LIBs. At the same time, it is substantially lower than the 1.5 V operating voltage of lithium titanate, thereby yielding a much higher energy density. 

Despite its great promise, there remains major ambiguity on the intercalation chemistry of DRX-\ce{Li3V2O5}. Previously, Zheng et al.\cite{zhang2017facile} proposed that the related DRX-\ce{Li_{0.78+x}V_{0.75}O2} anode undergoes a conversion-type reaction, in which the single phase material converts to \ce{VO2} and \ce{Li2O} as it is discharged to 0.55 V vs \ce{Li/Li^+}. However, previous density functional theory (DFT) calculations by the present authors attributed the low voltage and high rate capability of DRX-\ce{Li3V2O5} to a redistributive lithium intercalation mechanism with low energy barriers.\cite{liu2020disordered} These initial conclusions were reached based on 0K DFT calculations with small model cells, which did not fully explore the configurational space of the DRX anode at finite temperatures.

In this work, we revisit the intercalation chemistry of the DRX-\ce{Li_{3+x}V2O5} ($0 \leq x \leq 2)$) at finite temperatures using large-scale simulations with machine-learned energy models. Monte Carlo simulations using a fitted cluster expansion model predict that Li primarily inserts into the tetrahedral sites of DRX-\ce{Li3V2O5} while Li occupancy in octahedral sites remains largely unchanged, in contrast with earlier DFT results. Molecular dynamics (MD) simulations using a machine learning interatomic potential reveal that \ce{Li+} diffusivity reaches the maximum at intermediate states of charge, but sharply decreases at the start of charge/discharge. The exceptionally high rate capability and superior cycling stability of DRX-\ce{Li3V2O5} anode are a result of facile diffusion of Li ions through tetrahedral - octahedral - tetrahedral pathway, consistent with previous nudged elastic band calculations.

\section{Methods}

\subsection{Structure model}

Figure \ref{fig:pd_0K}(a) shows the crystal structure of DRX-\ce{Li_{3+x}V2O5}, which has an fcc lattice with spacegroup $Fm\bar{3}m$. The \ce{O^{2-}} anions occupy $4a$ sites. The initial composition of \ce{Li3V2O5} has a cation: anion ratio of 1:1. Based on our previous studies, the lowest DFT energy structure is one where all Li/V cations fully occupy the octahedral 4$b$ sites.\cite{liu2020disordered} Henceforth, we will use the anion-normalized composition, e.g., \ce{[Li_{0.6}V_{0.4}]^{oct}O} for \ce{Li3V2O5}, to emphasize the site occupancies. During discharge, inserted \ce{Li+} also occupy the tetrahedral 8$c$ sites, forming \ce{Li^{tet}_x[Li_yV_{0.4}]^{oct}O}.


For more costly DFT and climbing image nudged elastic band (NEB) calculations, a set of 64 special quasi-random structures (SQSs)\cite{zunger1990special} with formula \ce{Li_{19}V_{13}O_{32}} (corresponding to \ce{[Li_{0.59375}V_{0.40625}]^{oct}O}) were generated using $2\times 2 \times 2$ supercells of the conventional rocksalt cubic unit cell. The three relaxed SQS structures with the lowest energies were used for site energy and migration barrier calculations. 

\subsection{Density functional theory calculations}
Spin-polarized density functional theory (DFT) calculations were performed using the Vienna \textit{ab initio} simulation (VASP) package within the projected-augmented wave method.\cite{kresse_efficient_1996, blochl_projector_1994} The Perdew-Burke-Ernzerhof (PBE) functional\cite{perdew_generalized_1996} was used for the structural relaxation and electronic energy calculations with an effective Hubbard $U$\cite{dudarev_electron-energy-loss_1998, jain2011formation} value of 3.25 eV for V, which is in line with the parameters used in the Materials Project\cite{jain_commentary:_2013}. All calculations were initialized in a ferromagnetic high-spin configuration. A plane wave energy cutoff of 450 eV and a $k$-point density of at least 100 per reciprocal volume were adopted. The electronic energies and forces were converged to 10$^{-5}$ eV and 0.02 eV \r{A}$^{-1}$, respectively.

\subsection{Li site energies}

The Li site energies were determined by inserting one \ce{Li+} into each symmetrically distinct tetrahedral site of the three lowest energy SQSs. The \ce{Li^{tet}} site energies are given by the following expression:
\begin{equation}
    E_{Li^{tet}} = E_{Doped} - E_{Host} - E_{Li}^0
\end{equation}
where $E_{Doped}$, $E_{Host}$, and $E_{Li}^0$ are the DFT calculated energies of the SQS with one Li atom inserted into the tetrahedral site, the SQS and Li metal, respectively.

\section{Order parameter}

To quantify the degree of ordering, the average Steinhardt order parameter ($\bar{D_j}$)\cite{kawasaki2011construction} over all atoms in each structure was calculated. The Steinhardt order parameter $D_j$ for each atom j and the bounded atom k in the structure is given by:
\begin{equation}
    D_j = \frac{1}{n^j_b}\sum_{k\in bonded}[S_{ij} + S_{kk} - 2S_{jk}]
\end{equation}
\begin{equation}
     S_{jk} = \sum_{-l\leq m\leq l}q^j_{lm}(q^k_{lm})^*
\end{equation}
where $S_{jk}$ is the correlation between atoms $j$ and $k$, $n^j_b$ is the number of the atoms bonded to the atom $j$. For $fcc$ cation lattice, $n^j_b = 12$ and $l = 6$ is used. For a perfect ordered crystal such as the layered $R\bar{3}m$ \ce{LiVO2}, $\bar{D_j} = 0$. The higher the degree of disordering of a system, the larger the value of $\bar{D_j}$. For example in glass material, the $\bar{D_j}$ ranges from 0.2 $\sim$ 0.3.\cite{kawasaki2011construction} In this work, the $\bar{D_j}$ was calculated using the pyscal package.\cite{Menon2019}

\subsection{Cluster expansion}

A cluster expansion (CE) lattice model\cite{van2010linking, sanchez1984generalized,de1994cluster,thomas2013finite} for the DRX \ce{Li^{tet}_x[Li_yV_{0.4}]^{oct}O} system was parameterized using the Clusters Approach to Statistical Mechanics (CASM) software.\cite{casm, li1994lattice} The DFT calculated total energies of the structure configurations were mapped into an expansion of crystal basis functions $\Phi(\vec{m}, \bar{\sigma})$ as given by
\begin{equation}
E(\bar{\sigma}) = \sum_{\vec{m}}V(\vec{m})\Phi_{\alpha}(\vec{m}, \bar{\sigma})
\end{equation}
where $\Phi(\vec{m}, \bar{\sigma}) = \prod_{n=1}^{N}\phi(n, m_n, \sigma_n)$ is a polynomial of site basis functions $\phi(n, m_n, \sigma_n)$, and V($\vec{m}$) are the fitted effective cluster interactions (ECIs).

To fit the CE, the \ce{[Li/V/Va]^{oct}} and \ce{[Li/Va]^{tet}} configurations of different compositions were exhaustively enumerated in cubic rocksalt supercells up to a maximum cell size of 25 times the primitive unit cell. It should be noted that V is allowed to occupy only the 4b octahedral sites, given that its occupancy at tetrahedral sites is extremely energetically unfavorable ($E_{V^{tet}} \sim$ 3 eV). Li can occupy either the 4b octahedral or 8c tetrahedral sites. We note that only structures with basis deformation $<$ 0.1 were used in the fit, which is a typical threshold used to identify structures that match with the primitive unit cell. The basis deformation is determined by the mean-square atomic displacement relative to the positions of the ideal ions in cubic rocksalt lattice.\cite{casm}  In total, DFT computed energies of about 4500 symmetrically distinct configurations were used to fit the ECIs. All symmetrically distinct pairs, triplets, and quadruplets in the rocksalt cell within a radius of 7.1\r{A}, 4.1\r{A}~and 4.1 \r{A}, respectively, were used to construct the CE model. 

The ECIs were obtained from a $L_1$ regularized linear regression fit with $\alpha$ = $10^{-8}$ to minimize over-fitting (Figure S4). The cross-validation mean absolute error in energies is 8.94 meV atom$^{-1}$. 

The calculated formation energy of \ce{Li_{3+x}V2O5} structures is given by 
\begin{equation}
    E_{form}[\ce{Li_{3+x}V2O5}] = E[\ce{Li_{3+x}V2O5}] - (1-\frac{x}{2})E[\ce{Li3V2O5}] - \frac{x}{2}E[\ce{Li5V2O5}]
\end{equation}

\subsection{Monte Carlo simulations}

Monte Carlo (MC) simulations were performed using the fitted cluster expansion model. A 5$\times$5$\times$5 supercell (2000 sites)  was used for these simulations; larger supercells produced similar results (see Figure S5). Canonical MC simulations were performed at the \ce{[Li_{0.6}V_{0.4}]^{oct}O} composition to probe the temperature under which the system undergoes an order-disorder transition.

Semi-grand canonical MC (GCMC) simulations at 300K were performed to study the intercalation of Li ions into the DRX-\ce{Li3V2O5} structure. In a semi-grand canonical ensemble, the composition and energy of the system with a fixed number of sites were allowed to fluctuate while the temperature ($T$) and the chemical potentials of Li ($\mu_{Li}$) and V ($\mu_{V}$) were externally imposed. The chemical potentials were referenced to that of bulk Li and V metals, which have $\mu_{Li} = \mu_{V} = 0$. The semi-GCMC simulations were carried out by scanning $\mu_{Li}$ with a step of $\delta\mu_{Li}$ = 0.01 eV at constant $\mu_{V}$ within the chemical potential ranges of $ -2.0 \leq \mu_{Li} \leq 0.0$ and $ -2.0 \leq \mu_{V} \leq 0.0$ at 300 K. The initial disordered structures were obtained from equilibrated semi-GCMC simulations by heating the system from 5 K up to 1500 K at fixed chemical potentials. In this work, the semi-GCMC simulations were performed on only Li and vacancy occupancy in the tetrahedral and octahedral sites, while all V ions were fixed in the initial equilibrated octahedral sites due to the large V migration barriers.

The voltage of an electrochemical cell was related to the Li chemical potential of the electrodes according to the Nernst equation:
\begin{equation}
    V = -(\mu_{Li} - \mu_{Li}^0)/e
\end{equation}
where $\mu_{Li}$ is the chemical potential of Li in DRX-\ce{Li_{3+x}V2O5}, $\mu_{Li}^0$ is the reference chemical potential of Li metal and $e$ is the elementary charge.

\subsection{Moment tensor potential}

A machine learning interatomic potential (ML-IAP) based on the moment tensor potential (MTP) formalism\cite{novikov2020mlip} was developed for DRX-\ce{Li_{3+x}V2O5} using a protocol similar to that developed by Qi et al.\cite{qi2021bridging} for lithium superionic conductors. The initial training structures included DFT calculated ground state structures in the Li-V-O chemical space (supercells of \ce{Li2O}, \ce{Li3VO4}, \ce{LiV2O5} and \ce{LiVO2} with lattice parameters larger than 10 \r{A} ). To further sample the energy landscape of different Li/V arrangements, a set of SQSs with compositions \ce{Li_{19}V_{13}O_{32}} (\ce{[Li_{0.59375}V_{0.40625}]^{oct}O}) $\approx$ \ce{Li3V2O5}), \ce{Li_{25}V_{13}O_{32}}(\ce{[Li_{0.1875}]^{tet}[Li_{0.59375}V_{0.40625}]^{oct}O}) $\approx$ \ce{Li4V2O5}) and \ce{Li_{32}V_{13}O_{32}}(\ce{[Li_{0.40625}]^{tet}[Li_{0.59375}V_{0.40625}]^{oct}O}) $\approx$ \ce{Li5V2O5}) were generated in 2$\times$2$\times$2 supercells of the conventional rocksalt cubic unit cell. It should be noted that changes in the occupancy of tetrahedral and octahedral sites in SQSs can occur upon  DFT relaxation. The SQS with the lowest energy configuration of each composition were included in the fitting procedure.  

Non-spin polarized \textit{ab initio} molecular dynamics (AIMD) simulations using NVT ensembles were performed on relaxed supercells of all initial structures with a plane-wave energy cut-off of 280 eV and Gamma $k$-point. To diversify the sampled local environments, the simulations were carried out at three strains (0, $\pm$ 0.05) and four temperatures (300 K - 1200 K at an interval of 300 K). All simulations were performed for at least 30 ps with a 2 fs timestep using Nose-Hoover thermostat\cite{nose1984unified, hoover1985canonical}. The training structures were collected from 15 ps-equilibrated runs at 0.1 ps intervals, and more accurate energies and forces were obtained by static self-consistent calculations with a $k$-point density of at least 100 per reciprocal volume and an energy cutoff of 520 eV.

\subsection{Molecular dynamics simulations}

Molecular dynamic (MD) simulations were performed in the $NpT$ ensemble to investigate the \ce{Li+} diffusion properties of DRX-\ce{Li3V2O5} using the fitted MTP potential. The time step of all MD calculations was set to 1 fs, and the total simulation time was at least 1 ns. The diffusivity ($D_{Li^+}$) of \ce{Li+} was calculated by performing a linear fitting of the mean square displacement (MSD) versus $2dt$:
\begin{equation}
D = \frac{1}{2dt}<[\Delta r(t)]^2>
\end{equation}
where $d$ is the dimensionality factor, which equals 3 for bulk structures. $<[\Delta r(t)]^2>$ is the average MSD over a time duration $t$. The activation energy of $E_a$ was determined by the Arrhenius relationship, 
\begin{equation}
    D = D_0exp(-E_a/kT)
\end{equation}
where $D_0$ is the maximum diffusivity at infinite temperature, $k$ is the Boltzmann constant and T is the temperature.

All the training, evaluations and MD simulations were performed using MLIP\cite{shapeev2016moment,gubaev2019accelerating}, Large-scale Atomic/Molecular Massively Parallel Simulator (LAMMPS)\cite{thompson2022lammps} and the open-source Materials Machine Learning (maml) Python package\cite{maml}.

\subsection{Diffusion barriers}

The migration barriers of \ce{Li+} and \ce{V^{n+}} (n=3, 4) vacancy were calculated using the climbing image nudged elastic band (CI-NEB) method\cite{henkelman_climbing_2000, henkelman_improved_2000}. All NEB calculations were performed in three SQSs with the lowest DFT total energy created in the previous sections. For \ce{Li+} migration barriers, the starting point of each NEB calculation was determined by inserting one \ce{Li+} into the tetrahedral site, and then the structure was fully relaxed. Five linearly interpolated intermediate images were used to generate the initial guess for the minimum energy path. For \ce{Li_{tet}} in varied local environments, the energy barriers for the \ce{Li+} hoping were calculated in three configurations with the lowest \ce{Li_{tet}} site energy. The kinetically resolved activation (KRA)\cite{van2001first} \ce{Li+} migration barrier ($\Delta E_{KRA}$), which is independent of hop direction, was determined by the following expression:
\begin{equation}
    \Delta E_{KRA} = E(\sigma_t) - \frac{1}{2}(E(\sigma_i + E(\sigma_f))
\end{equation}
where $E(\sigma_t)$, $E(\sigma_i$ and $E(\sigma_f)$ refer to the energy of the activated transition state, the initial state and the end state from the CI-NEB calculations, respectively. The PBE functional without Hubbard $U$ was adopted to avoid the possible mixing of the diffusion barrier with a charge transfer barrier. \cite{ong_voltage_2011} The force and energy convergence criterion was 0.05 eV/\r{A} and 5$\times$10$^{-5}$ eV. 

\section{Results}

\subsection{Li-V-O phase diagram}

\begin{figure}[htp!]
\centering
\includegraphics[width=1.0\textwidth]{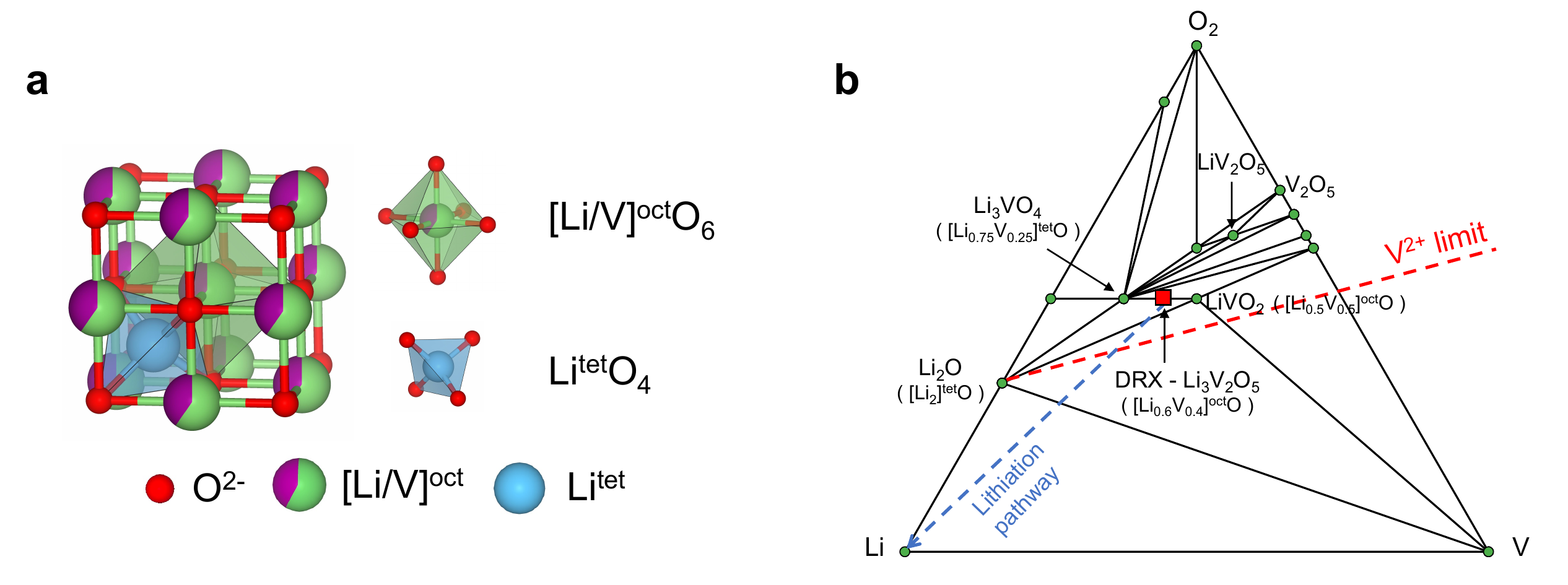}
\caption{\label{fig:pd_0K} (a) The crystal structure of disordered rocksalt \ce{Li_{3+x}V2O5}. Red: \ce{O^{2-}} anions forming an fcc sublattice. Light blue: \ce{Li+} in tetrahedral interstitials. Green/purple: Disordered \ce{Li+}/\ce{V^{2+-4+}} in octahedral interstitials. (b) Calculated phase diagram of Li-V-O chemical system at 0 K. The green circles refer to ground states and the red square refers to the metastable phase. The red dashed line refers to the limit of \ce{V^{2+}} oxidation state. The blue dashed line refers to the lithiation pathway in DRX-\ce{Li3V2O5}.}
\end{figure}

Figure \ref{fig:pd_0K}(b) shows the DFT-calculated Li-V-O phase diagram at 0 K. \ce{Li3V2O5} (\ce{[Li_{0.6}V_{0.4}]^{oct}O}), represented by the red square, is unstable at 0K relative to \ce{LiVO2} and \ce{Li3VO4}. Both ordered layered and DRX-\ce{LiVO2} as well as ordered \ce{Li3VO4} have been explored as potential LIB anodes\cite{armstrong2011lithium,li2013li3vo4,baur2018reversible}. All three electrode materials have a cation: anion ratio of 1:1 and differ purely in their Li: V ratio and therefore active redox couple. \ce{Li3VO4} has a formal vanadium oxidation state of 5+, and all cations are ordered in tetrahedral sites. It can therefore be represented using the anion-normalized composition of \ce{[Li_{0.75}V_{0.25}]^{tet}O}. \ce{LiVO2}, on the other hand, has full octahedral occupancy like \ce{Li3V2O5} and a formal V oxidation state of 3+. It can therefore be represented using the anion-normalized composition of \ce{[Li_{0.5}V_{0.5}]^{oct}O}. It should be noted that \ce{Li2O} (\ce{[Li_2]^{tet}O}) also has a cubic rocksalt structure with full occupancy in the tetrahedral sites and a cation: anion ratio of 2.

During charging, lithiation of \ce{[Li_{0.6}V_{0.4}]^{oct}O} occurs along the pathway indicated by the dashed blue line. The thermodynamically most favorable pathway is a conversion reaction given by the phase triangle of \ce{Li2O}-\ce{Li3VO4}-\ce{LiVO2}(\ce{[Li_2]^{tet}O}-\ce{[Li_{0.75}V_{0.25}]^{tet}O}-\ce{[Li_{0.5}V_{0.5}]^{oct}O}). However, kinetic considerations may favor lithium insertion instead of conversion, as has been observed experimentally.\cite{liu2020disordered} The ultimate limit of insertion into the rock salt \ce{Li_{1-x}V_{x}O} is given by the red dashed line representing \ce{Li_{2-2x}V_{x}O}, i.e., a \ce{V^{2+}} oxidation state. For a V cation content of 0.4, i.e., \ce{Li3V2O5}, this limit is given by \ce{Li_{1.2}V_{0.4}O} or \ce{Li6V2O5}. Experimentally, this limit is never reached and the highest lithiated state has a composition of \ce{Li_{1}V_{0.4}O} or \ce{Li5V2O5}.\cite{liu2020disordered}

\begin{figure}[htp!]
\centering
\includegraphics[width=1.0\textwidth]{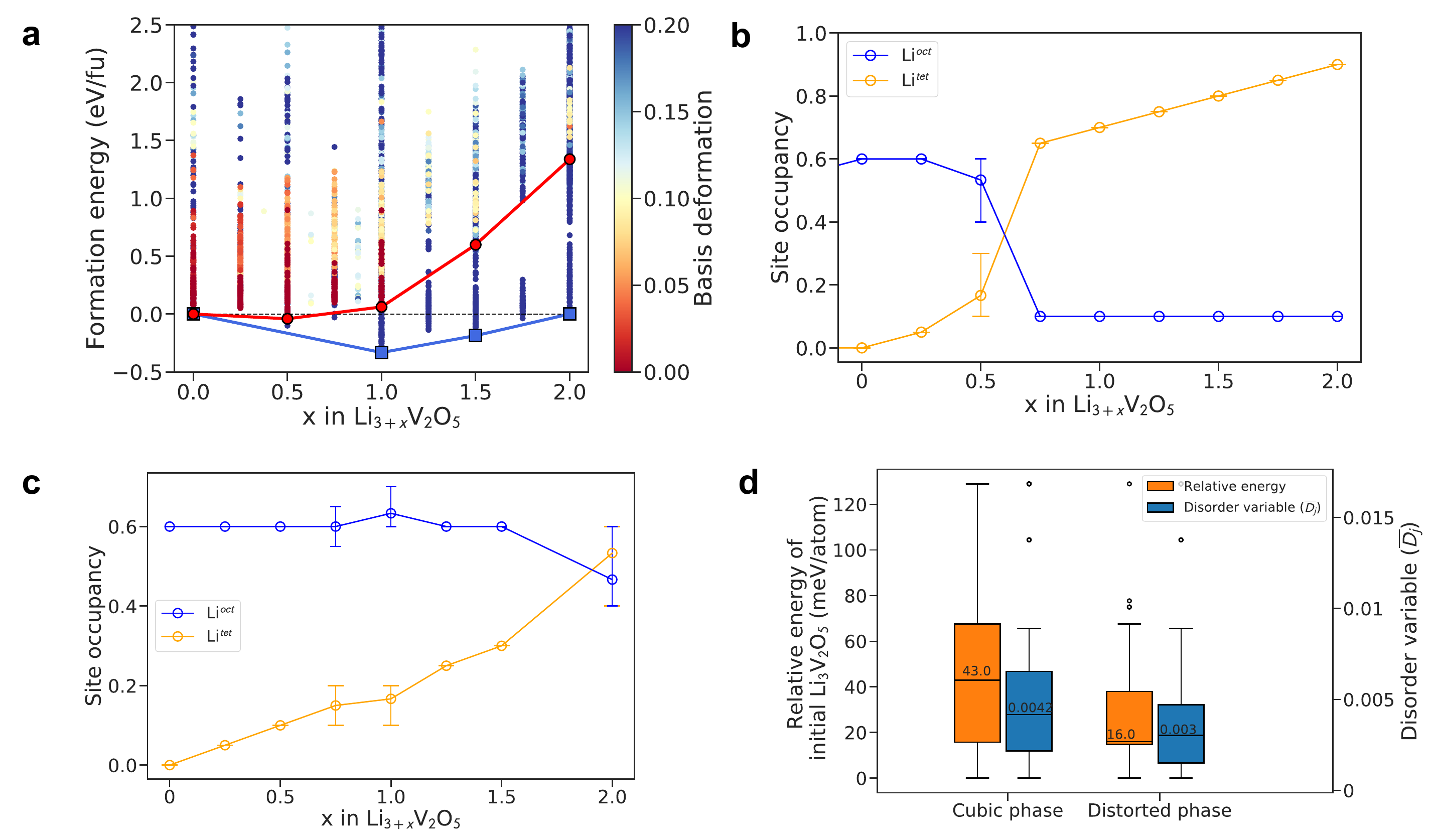}
\caption{\label{fig:site_occu_dis} (a) DFT calculated pseudo-binary \ce{Li3V2O5}-\ce{Li5V2O5} compound phase diagram. The configurations are colored in terms of their basis deformation. The red line represents the convex hull of the cubic phase, i.e., structures with low basis deformation ($<$ 0.1). The blue line represents the convex hull of all the DFT relaxed structures. (b) Evolution of the occupancy of Li$^{oct}$ and Li$^{tet}$ within \ce{Li3V2O5} - \ce{Li5V2O5} at 0 K. (c) Evolution of the occupancy of Li$^{oct}$ and Li$^{tet}$ within \ce{Li3V2O5} - \ce{Li5V2O5} in cubic phase. At each composition, the three lowest-energy structures were used for site occupancy calculations. The average values with error bars are shown in the plot. The most stable structures of ground states and metastable cubic phase of \ce{Li_{3+x}V2O5} (x = 0, 1, 2) are shown in Figure S3. (d) The distribution in DFT relative energies and average Steinhardt order parameter $\bar{D_j}$ of the initial \ce{Li3V2O5} structures corresponding to the cubic and distorted lithiated \ce{Li_{3+x}V2O5} ($x \geq 1$) structures. For each composition, no more than 50 structures with the lowest energies were used for the analysis. }
\end{figure}

\ce{Li+} insertion into the \ce{Li3V2O5} anode at 0 K were studied by DFT calculations. Figure \ref{fig:site_occu_dis}(a) shows the DFT calculated pseudo-binary \ce{Li3V2O5}-\ce{Li5V2O5} phase diagram with the candidate structures colored in terms of their basis deformation. The larger the basis deformation, the greater the deviation of the DFT-relaxed structure from the ideal cubic rocksalt lattice. At $x = 0$, low energy structures have low basis deformation and Li/V atoms are all in octahedral sites. At $x \geq 0.5$, the DFT ground states comprise phases that are highly distorted (basis deformation $>$ 0.1) from the parent cubic rocksalt lattice; the cubic phases (basis deformation $\leq$ 0.1) are metastable. A Li site occupancy analysis of DFT ground states structures (Figure \ref{fig:site_occu_dis}(b)) indicates that redistribution of Li from octahedral sites to tetrahedral sites occurs at x $\sim$ 0.75, where the occupancy of Li$^{oct}$ sharply decreases to 0.1 and the occupancy of Li$^{tet}$ increases to 0.65. These results are qualitatively in line with previous DFT studies.\cite{liu2020disordered} However, this redistribution of Li is accompanied by a considerably large volume change ($\sim 20\%$) and lattice distortion, contradicting with the experimental observations that the anode retains a stable cubic lattice with a small volume change of 5.9\%.\cite{liu2020disordered} In contrast, the Li site occupancy analysis of metastable cubic phases (Figure \ref{fig:site_occu_dis} (c)) suggests that a redistribution of Li only takes place at the end of charging. The volume change of these metastable cubic phases ranges from 8 to 15\%, which is close to the experimental results.\cite{liu2020disordered}. 

Figure \ref{fig:site_occu_dis}(d) shows the distribution in relative energies and the average Steinhardt order parameter ($\bar{D_j}$) of the initial \ce{Li3V2O5} structures corresponding to the cubic and distorted lithiated \ce{Li_{$3+x$}V2O5} ($x \geq 1$) structures. It can be seen that the lithiated structures that maintain a cubic-like structure have greater $\bar{D_j}$, i.e., more Li/V disorder, and higher relative energies than those that are highly distorted (see Figure S2(b)). These cubic-like structures also exhibit only moderate Li redistribution at the end of charge. In contrast, the low DFT energy lithiated structures correspond to initial Li/V arrangements that are less disordered. In fact, the DFT ground state \ce{Li3V2O5} exhibit an ordered layered-like Li/V arrangement akin to \ce{LiVO2} with $\bar{D_j} = 6\times 10^{-4}$. 

It should be noted that the initial Li/V disorder is controlled via synthesis levers such as temperature. Given that the as-synthesized DRX-\ce{Li3V2O5} anode is highly cubic and exhibits Li/V disorder, we may surmise from the above results that the relevant lithiation pathway is one in which Li primarily inserts into the tetrahedral sites with minimal redistribution of Li$^{oct}$ to Li$^{tet}$ while maintaining a cubic-like framework with low distortion. 

\subsection{Validation of cluster expansion model}

\begin{figure}[htp!]
\centering
\includegraphics[width=1.0\textwidth]{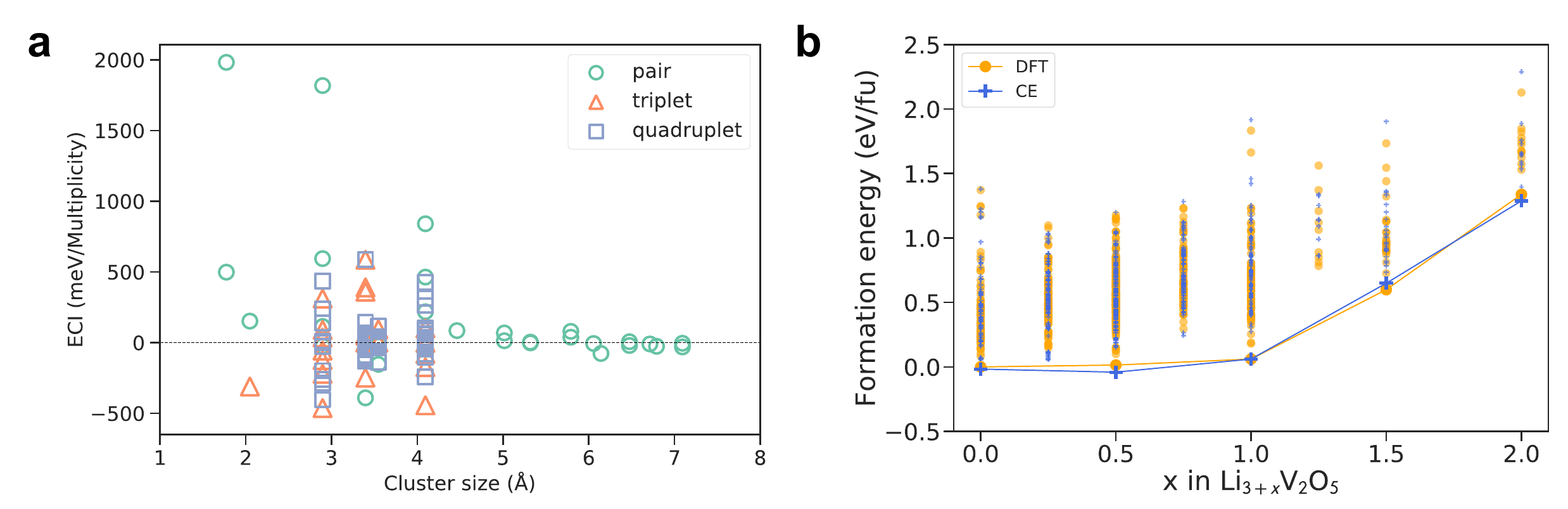}
\caption{\label{fig:eci_size} (a) Fitted effective cluster interactions (ECI) values with respect to the size of the clusters (b) DFT calculated and cluster expansion (CE) predicted formation energy of \ce{Li_{3+x}V2O5} (x = 0.0 - 2.0)}
\end{figure}

Figure\ref{fig:eci_size} (a) plots the 112 non-zero ECIs of the fitted CE model versus cluster size. It may be observed that the magnitude of the ECIs converge to near 0 at the cutoff radius. From Figure\ref{fig:eci_size} (b), it may be observed that the fitted CE model accurately captures the DFT computed ground states (structures with formation energy on the convex hull).

\subsection{Order-disorder transition of cubic \ce{Li3V2O5}}
\begin{figure}[htp!]
\centering
\includegraphics[width=1.0\textwidth]{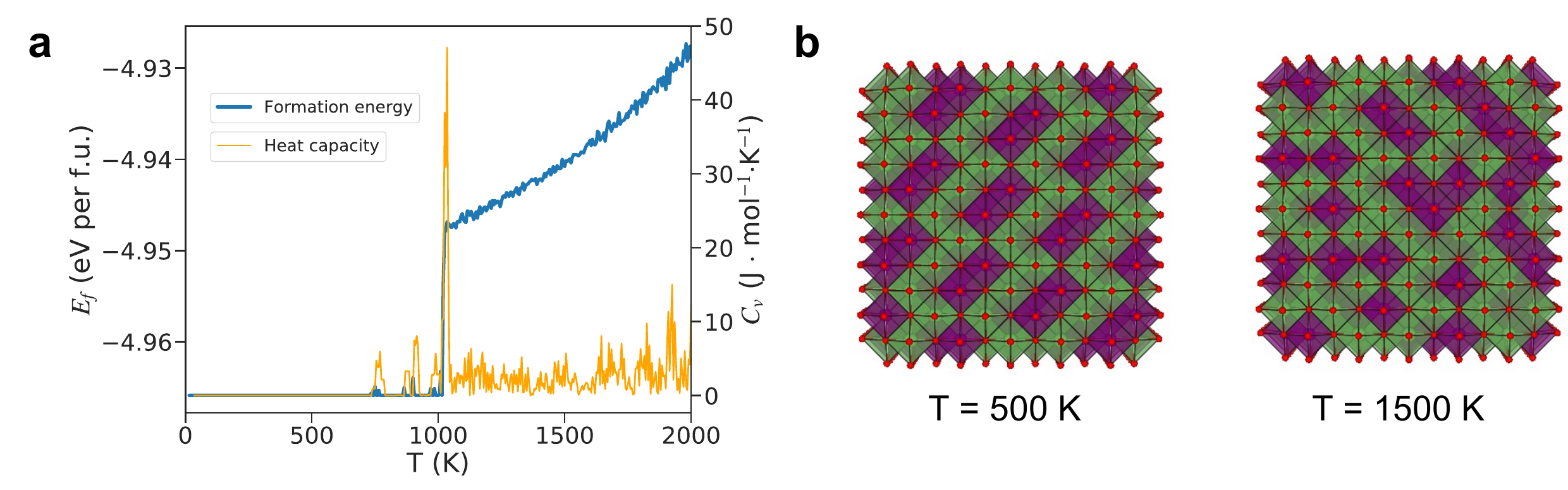}
\caption{\label{fig:order_disorder} (a) Calculated formation energy ($E_f$) and heat capacity ($C_v$) as a function of temperature ($T$). Structures from equilibrated Monte Carlo simulations at (b) 500 K and 1500 K. The MC simulations were initialized in the DFT calculated most energetically stable \ce{Li3V2O5} structure, in which all Li and V ions occupy the octahedral sites. The initial configuration was then heated from 10 K to 2000K in intervals of = 10 K. At each temperature, the properties were then obtained by averaging the results from 1000 equilibrated MC runs. The configurational heat capacity $C_v$ is given by the second derivative of the formation energy $E_f$ with respect to temperature $T$, $C_v = \partial^2 E_f/\partial^2 T$.}
\end{figure}

To further understand the effect of cation disorder in DRX-\ce{Li3V2O5} anode, we performed Monte Carlo simulation using the fitted CE model. Figure \ref{fig:order_disorder} shows the calculated formation energy $E_f$ and specific heat capacity $C_v$ as a function of temperature from MC simulations with the fitted CE model. The critical temperature ($T_c$) of order-disorder phase transition is predicted to be $\sim$ 1000 K, which is somewhat lower than that experimentally measured from \ce{LiVO2} (1775 K)\cite{hewston1987survey}. At temperatures below $T_c$, both $E_f$ and $C_v$  remain relatively constant with temperature, characteristic of a highly ordered crystal. At $T_c$, a discontinuous increase in the $E_f$ and a sharp peak in the configurational $C_v$ is observed, indicative of a phase transition taking place. The calculated average configurational specific heat capacity of the DRX phase ( T $>$ 1000 K) is 2.973 J mol$^{-1}$ K$^{-1}$. Figure \ref{fig:order_disorder} (b) shows two sampled configurations from the equilibrated MC simulations at 500K (ordered phase) and 1500 K (fully disordered phase). At temperatures above $T_c$, Li/V disordering occurs mostly on the octahedral 4$b$ sites and only a small fraction of Li ($\sim$ 0.0125 $\%$) ions occupy the tetrahedral sites. These observations are consistent with the fact that the DRX is synthesized under the application of an external driving force, such as high temperatures, ball milling, or electrochemical lithiation.\cite{delmas1995formation, liu2020disordered, barnes2022electrochemically}

\subsection{Lithium intercalation mechanism and predicted voltage profile}
\begin{figure}[htp!]
\centering
\includegraphics[width=1.0\textwidth]{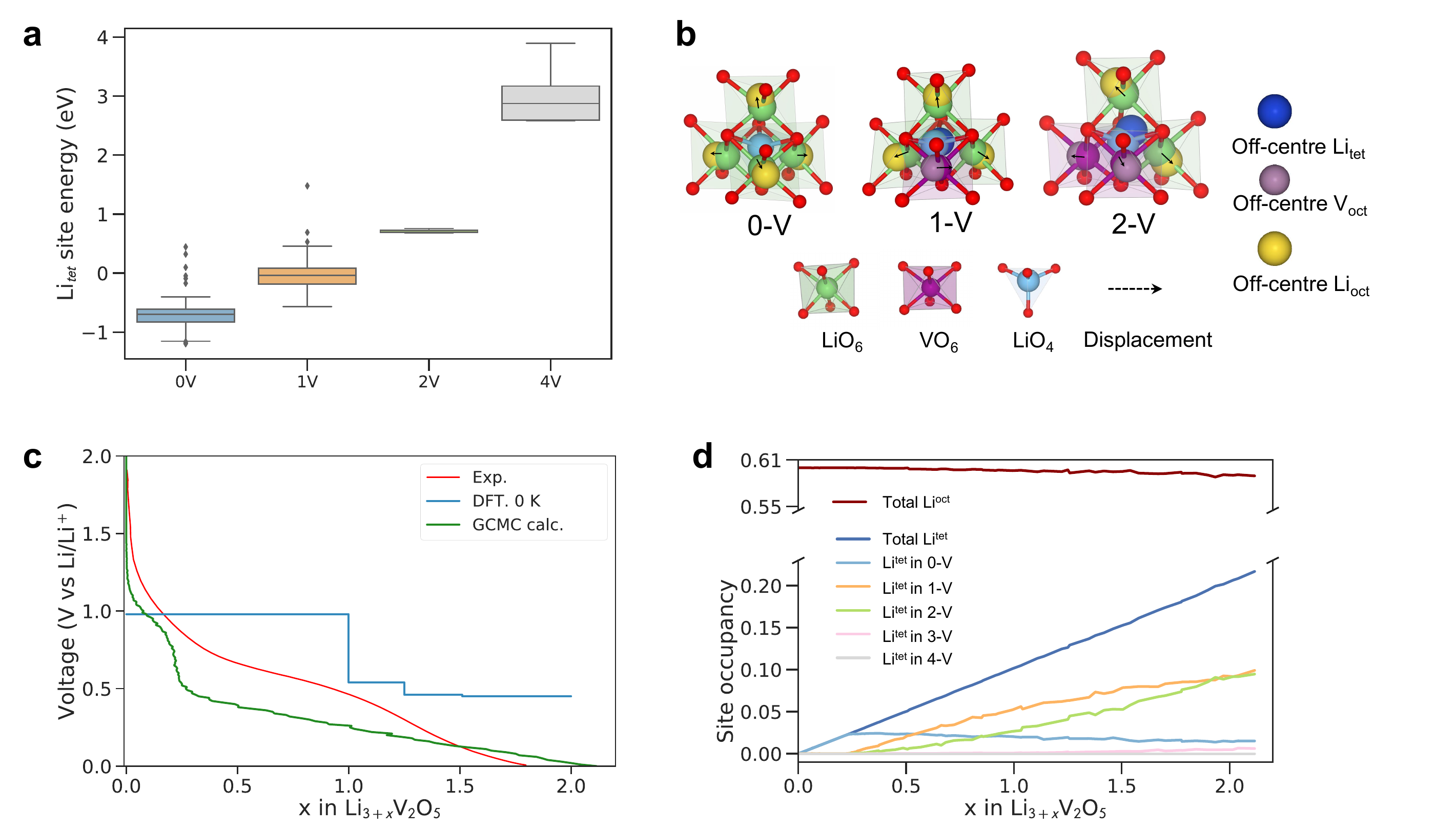}
\caption{\label{fig:local_env_sqs}(a) Site energy of Li$^{tet}$ in various local environments (i.e., diffusion sites). (b) 0-V, 1-V, and 2-V tetrahedral Li insertion sites and off-center displacements of the neighboring octahedral sites and the inserted tetrahedral Li sites. (c) Calculated voltage as a function of Li content x in \ce{Li_{3+x}V_2O_5}. The experimental and DFT (0 K) results are extracted from \citet{liu2020disordered} (d) Evolution of Li site occupancies in tetrahedral and octahedral sites upon Li insertion from semi-GCMC simulations with the position of $V^{oct}$ fixed.}
\end{figure}

The intercalation mechanism of \ce{Li+} into DRX structures is strongly affected by the distribution of the tetrahedral sites with different local environments and connectivity. The Li in tetrahedral sites share faces with four neighboring octahedral sites. As shown in previous works,\cite{lee2014unlocking,urban2016computational,liu2020disordered,clement2020cation} the Li/transition metal occupancy of these neighboring octahedral sites have a strong influence on the local site occupancies and \ce{Li+} migration barriers. The local environment of a tetrahedral site can be denoted by $n$-V ($n$ = 0 $\sim$ 4), where $n$ represents the number of face-shared V atoms. Figure\ref{fig:local_env_sqs} (a) shows the distribution of the calculated \ce{Li^{tet}} site energies in 0-V, 1-V, 2-V and 4-V clusters in the three SQSs with the lowest DFT energies. The \ce{Li_{tet}} site energy in 3-V is not presented because \ce{Li_{tet}} is unstable in this local environment and the electrostatic repulsive effect between the inserted \ce{Li_{tet}} and the neighboring \ce{Li^{oct}} ion caused local environment rearrangements during structure relaxations. In general, the \ce{Li_{tet}} site energy and the distortion of neighboring \ce{LiO6} octahedra increase with the number of neighboring V atoms. The Li insertion into 0-V sites is the most energetically favourable and the four neighboring \ce{Li_{oct}} sites are displaced off-center by 0.4-0.5 \r{A} due to the electrostatic repulsion between \ce{Li+} (Figure \ref{fig:local_env_sqs}(b)). The 1-V \ce{Li_{tet}} site energy is on average 660 meV higher than that of 0-V sites and the inserted Li displaces $\sim$ 0.4 \r{A} from the center of the tetrahedron in the direction away from the V ions. The off-center displacements of neighboring \ce{Li^{oct}} are 0.3-0.6 \r{A}, whereas that for the neighboring \ce{V^{oct}} are only 0.16 \r{A}. The 2-V \ce{Li^{tet}} sites are relatively energetically unstable and is positioned $\sim$ 0.4 \r{A} away from the tetrahedral center due to large repulsive interactions between \ce{V^{3+}}/\ce{V^{4+}} and \ce{Li+}. The neighboring Li atoms are pushed 0.5 - 0.6 \r{A} away from the center of the octahedron, causing large distortions in the local lattice structure. 

The migration of V vacancies between neighboring octahedral sites was also investigated by introducing one $V^{oct}$ vacancy in the configuration with the lowest $V^{oct}$ vacancy energy. The average $V^{oct}$ migration barrier is extremely high, up to 2100 meV (Figure S1). This suggests that V ions in DRX-\ce{Li3V2O5} are unlikely to migrate during charge/discharge under operating conditions.

Figure \ref{fig:local_env_sqs} (c) shows the calculated voltage profile by GCMC simulations. The GCMC-predicted voltage profile exhibits a solid solution-like behavior and is in good agreement with the experimentally-measured voltage profile.\cite{liu2020disordered}  Li starts to insert into DRX-\ce{Li3V2O5} at $\sim$1.43 V and the predicted voltage profile exhibits two voltage steps at around 0.9 and 0.25 V vs Li/Li$^+$. As the voltage decreases to 0.01 V, the predicted average composition of the lithiated anode is \ce{Li_{5.11}V2O5}, close to that observed in experiments (\ce{Li_{4.86}V2O5})\cite{liu2020disordered}. In contrast, due to the limited supercell sizes used, DFT calculations\cite{liu2020disordered} predict a multiple voltage steps at 0.98V, 0.54 V, 0.46 V and 0.45 V.

Figure \ref{fig:local_env_sqs} (d) shows the evolution of Li site occupancies in DRX-\ce{Li_{3+x}V2O5} as a function of inserted Li content $x$. It should be noted that as an anode, insertion of Li into the DRX-\ce{Li3V2O5} corresponds to the charging process in a typical Li-ion battery. As Li ions are introduced into the tetrahedral sites, the Li occupancy in the octahedral sites first remains constant at around 0.6 until $x \approx 0.3$, following which there is a small, gradual reduction in Li$^{oct}$ occupancy. The insertion of Li ions into the tetrahedral sites consists of two steps. \ce{Li+} first intercalate into 0-V sites, which is consistent with the energies predicted by DFT calculations (Figure\ref{fig:local_env_sqs} (a)) and previous results that the insertion into 0-V tetrahedral sites is the most energetically favorable.\cite{liu2020disordered} This process coincides with the formation of the first plateau at around 0.9 V on the voltage profile. At x = 0.31, 0-V sites are fully occupied and further lithiation insertion occurs in the energetically less favorable 1-V and 2-V sites. The occupancy of 1-V and 2-V sites rise to $\sim$ 0.1 till the end of charge. At the end of charge, only $\sim$0.1$\%$ 3-V tetrahedral sites are occupied, and all 4-V sites are vacant. 

\subsection{Validation of moment tensor potential}

\begin{figure}[htp!]
\centering
\includegraphics[width=0.8\textwidth]{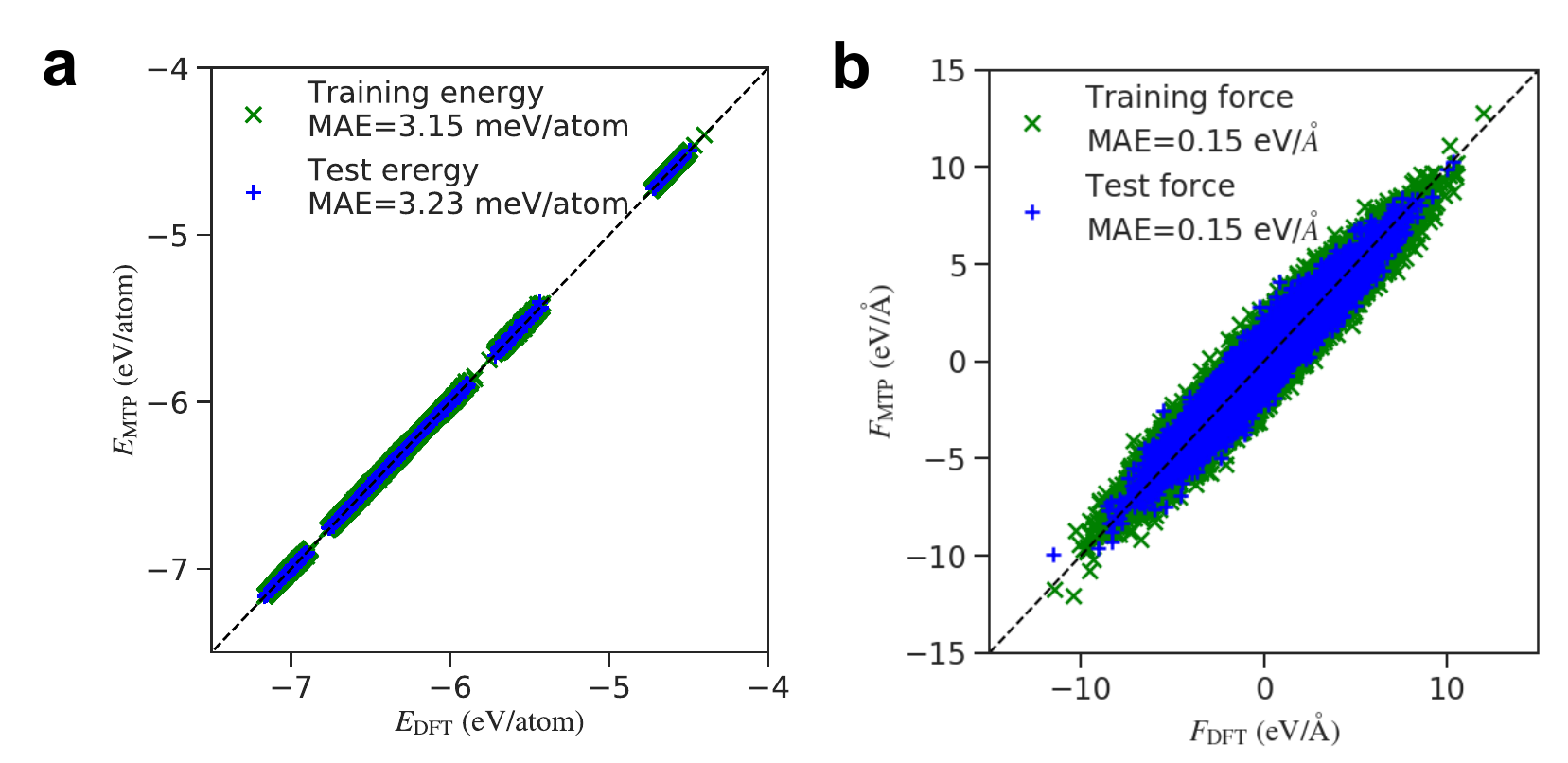}
\caption{\label{fig:parity_plot} Plots of the (a) MTP predicted versus DFT energies and (b)  MTP predicted versus DFT forces with $lev_{max}$ = 20.}
\end{figure}

The mean absolute errors (MAEs) of training energies and forces (shown in Figure \ref{fig:parity_plot}) of the fitted MTP are 3.15 meV/atom and 0.15 eV/\r{A}, respectively, comparable to those of other MTPs in the literature.\cite{zuo2020performance, wang2020lithium} As shown in Table \ref{table:mtp_lattice}, the MTP is able to accurately reproduce the lattice parameters and densities of training structures, with errors of within $\pm$ 3.12\% and $\pm$ 4.42\% relative to DFT values, respectively.

\begin{table}
\centering
\caption{\label{table:mtp_lattice} Lattice parameters and densities of the structures in the training set (zero strain) relaxed with the trained MTP at 0K, in comparison with DFT calculated lattice parameters and densities at 0K. Values in brackets are the percentage differences between the MTP and DFT computed values, respectively. }
\begin{tabular}{ccccc}
\hline
Composition & a (\r{A})            & b (\r{A})            & c (\r{A})           & Density (g/cm$^3$) \\\hline
\ce{LiVO2}       & 3.02 (1.75 \%)   & 5.16 (-1.24 \%)   & 5.25  (1.19 \%) & 3.88 (1.69 \%) \\
\ce{LiV2O5}      & 6.75 (0.98 \%)   & 7.68 (0.41 \%)   & 7.58 (-1.65 \%) & 3.18 (-0.28 \%) \\
\ce{Li3VO4}      & 5.06 (0.48 \%)  & 5.54 (0.51 \%)  & 6.37 (-0.41 \%) & 2.52 (-0.57 \%)  \\
\ce{Li2O}        & 3.30 (0.056 \%) & 3.29 (0.056 \%) & 3.29 (0.056\%) & 1.96 (-0.17\%) \\
DRX-\ce{Li3V2O5}   & 8.46 (0.66 \%) & 8.45 (1.26 \%) & 8.42 (0.84 \%)  & 3.60 (-2.71 \%) \\
DRX-\ce{Li_{3.9}V2O5} & 8.62 (1.26 \%)  & 8.62 (1.25 \%)  & 8.62 (1.24 \%) & 3.18 (-3.66 \%) \\
DRX-\ce{Li5V2O5}   & 9.48 (3.12 \%) & 7.72 (-1.03 \%) & 9.43 (2.519 \%) & 3.36 (-4.42 \%) \\ \hline
                                 
\end{tabular}
\end{table}

\subsection{Diffusion properties}
\begin{figure}[htp!]
\centering
\includegraphics[width=1.0\textwidth]{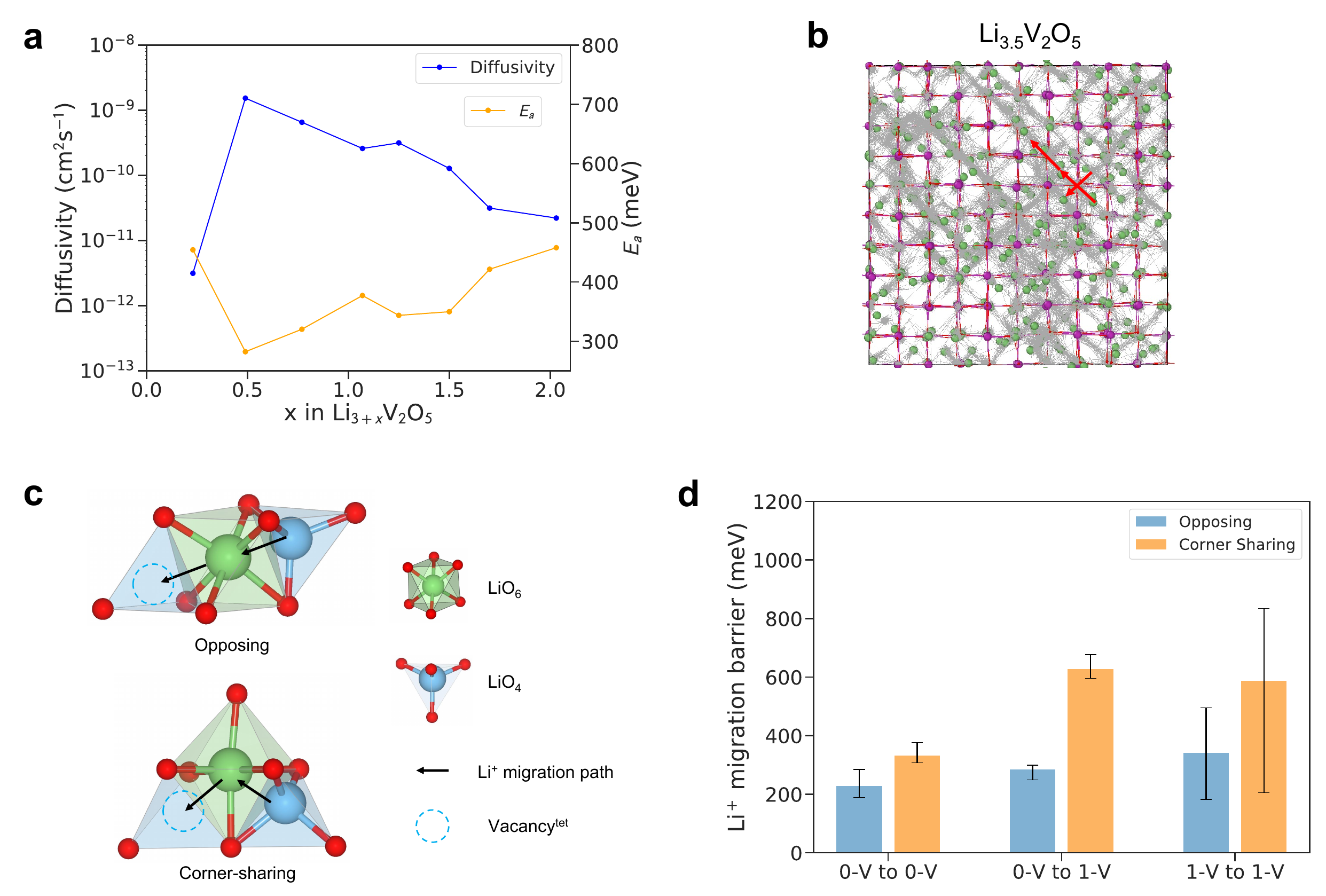}
\caption{\label{fig:diffusivity} (a) Calculated diffusivity and activation energy of \ce{Li+} in DRX-\ce{Li_{3+x}V2O5} as a function of Li content at 300 K. (b) Li trajectories (colored as grey) from MD simulations of \ce{Li_{3.5}V2O5} at 600 K, projected in the crystallographic a-b planes. Illustrations of ``t-o-t'' migration mechanisms of \ce{Li+} are shown in red arrows. The green balls, purple balls, and red balls represent Li, V, and O atoms, respectively.) (c) Illustration of \ce{Li+} migration path in DRX-\ce{Li3V2O5}. The opposing and corner-sharing pathways indicate cooperative migration mechanisms of \ce{Li+} via the octahedral site and its next tetrahedral site. (d) Calculated NEB barriers for possible Li migration hops. The barriers are categorized in terms of their mechanisms and local environments.}
\end{figure}

MD simulations were carried out to investigate the \ce{Li+} diffusion in the DRX-\ce{Li_{3+x}V2O5} anode as a function of Li concentration using the fitted MTP potential. The initial structures were obtained from the equilibrium semi-GCMC simulations in the preceding section. Figure\ref{fig:diffusivity} shows the calculated diffusivity $D_{Li^+}$ and activation energy $E_a$ as a function of Li content $x$ in DRX-\ce{Li_{3+x}V2O5} at room temperature (300 K). The Arrhenius plot from $NpT$ MD simulations for each composition is shown in Figure S6. At the start of lithiation, a sharp increase in $D_{Li^+}$ to $\sim 10^{-9}$ cm$^2$ s$^{-1}$ with a corresponding decrease in $E_a$ from 450 meV to 280 meV is observed up to $x = 0.5$. However, further lithiation results in a gradual decrease in $D_{Li^+}$ from $\sim 10^{-9}$ cm$^2$ s$^{-1}$ to $\sim 10^{-11}$ cm$^2$ s$^{-1}$ accompanied by an increase in $E_a$ from 280 to 460 meV.

We further analyzed the \ce{Li+} trajectories from 1 ns MD simulations at 600 K and the results for selected configurations are presented in Figure \ref{fig:diffusivity} (b). For $x < 1.5$, \ce{Li+} ions migrate via a cooperative tetrahedral - octahedral - tetrahedral (``t-o-t'') mechanism, in agreement with previous NEB calculations.\cite{liu2020disordered} For  $x > 1.5$, the increased amount of inserted \ce{Li+} in tetrahedral sites and vacancies in octahedral sites may result in \ce{Li+} migration between neighboring tetrahedral - tetrahedral (``t-t'') sites via the vacant octahedral sites along with the ``t-o-t'' migration.

To confirm the above results, DFT NEB calculations were also performed to calculate the \ce{Li+} migration barriers. Similar to the previous NEB calculations,\cite{liu2020disordered} a cooperative mechanism of \ce{Li+} migration was considered, where the tetrahedral \ce{Li+} migrates to a neighboring octahedral site and the octahedral \ce{Li+} migrates into another neighboring tetrahedral site (Figure \ref{fig:diffusivity} (c)). Here, we consider cooperative migration that occurs between corner-sharing and opposing tetrahedral sites, given that migration between edge-sharing tetrahedral sites is extremely energetically unfavorable due to the close proximity of the Li during migration (Figure S8(c) and S9(c)). Figure\ref{fig:diffusivity}(d) shows the calculated \ce{Li+} migration barriers between connected 0-V/1-V sites. We note that migration from 2-V site to 0-V and 1-V sites are energetically downhill due to the large site energy differences (Figure S11). The cooperative migration of \ce{Li+} via opposing ``t-o-t'' paths exhibits the lowest calculated average energy barriers, ranging from 230 to 340 meV in 0-V and 1-V sites. The average energy barriers of \ce{Li+} transport through corner-sharing 't-o-t' pathways exhibit higher energy barriers of 334 to 628 meV. In general, the migration barriers of \ce{Li+} increase as more V atoms gather around the local environment of tetrahedral \ce{Li+} sites. At the end of discharge, the formation of vacancies at the octahedral sites allows for direct hopping of \ce{Li+} from one tetrahedral site to its next edge-sharing tetrahedral site ('t - t'), as illustrated in Figure S13, with a low energy barrier of 241 meV. These results are in line with the activation energies obtained by MD simulations and previous theoretical studies\cite{liu2020disordered} that the facile migration of \ce{Li+} across opposing 't-o-t' pathways results in high rate capability of DRX-\ce{Li_{3+x}V2O5} anode. In previous studies\cite{liu2020disordered}, \ce{Li+} hopping via 0-V sites is the only mechanism that was considered while our results show that \ce{Li+} migration in 0-V, 1-V and 2-V sites is facile. Similar results have also been observed in LTO anodes that the transportation of \ce{Li+} in face-sharing octahedral-tetrahedral motif contributes to fast kinetics.\cite{zhang2020kinetic}

\section{Conclusions}

A major conclusion from this work is that the lithium insertion mechanism in \ce{Li3V2O5} depends greatly on the initial Li/V disorder. For highly ordered Li/V arrangements with lower energies, Li insertion occurs via a redistribution mechanism with the formation of highly distorted structures and accompanied by large volume changes. This is consistent with previous DFT results, but inconsistent with experimental observations for the DRX-\ce{Li3V2O5} electrode.\cite{liu2020disordered} In contrast, Li/V disorder leads to Li inserting primarily into the tetrahedral sites, while retaining a cubic-like structure with small volume changes. The MC-predicted voltage profile based on this latter mechanism is also in much better agreement with the experimental voltage profile. 

The \ce{Li+} insertion mechanism and transport depends on the local environment around each tetrahedral site and the distribution of sites. Our results show that \ce{Li+} in 0-V sites are the most energetically favorable and have the lowest migration barrier and $\geq$2-V sites are highly energetically unfavorable. Assuming a fully cation-disordered DRX-\ce{Li3V2O5} (\ce{[Li_{0.6}V_{0.4}]^{oct}O}), the probability of a $n$-V site $P(n)$ is given by the binomial distribution:
\begin{equation}
    P(n) = \frac{4!}{n!(4-n)!}0.4^{n}0.6^{4-n} 
\end{equation}

The fractions of 0-V, 1-V and $\geq$2-V are therefore 0.13, 0.35, and 0.52, respectively. The highest experimentally-realized lithiation state is \ce{Li5V2O5} ($\sim$ \ce{Li1V_{0.4}O}), which approximately corresponds to the full occupancy of the most energetically favored 0-V sites and about 0.77 occupancy of the 1-V tetrahedral sites. Hence, we can surmise that the inability of DRX-\ce{Li_{3+x}V2O5} to reach the theoretical \ce{V^{2+}} limit of \ce{Li6V2O5} is due to the lack of low energy sites for further Li insertion.

The above analysis suggest that one potential avenue to tune the electrochemical properties of the DRX anode is by varying the Li:V ratio in \ce{[Li_{1-x}V_{x}]^{oct}O}. Increasing the Li:V ratio would increase the fraction of 0-V/1-V tetrahedral sites, but at the price of increasing the V oxidation state and hence, the anode voltage. Conversely, lowering the Li:V ratio would decrease lithiation capacity but also decrease the anode voltage. The optimal ratio would depend on the specific application, for example, whether a high rate or a high energy density anode is prioritized.

\begin{acknowledgement}

This work is supported by the Materials Project, funded by the U.S. Department of Energy, Office of Science, Office of Basic Energy Sciences, Materials Sciences and Engineering Division under Contract no. DEAC02-05-CH11231: Materials Project program KC23MP. The authors acknowledge the computing resources provided by Triton Shared Computing Cluster (TSCC) at UC San Diego, the National Energy Research Scientific Computing Center (NERSC) and the Extreme Science and Engineering Discovery Environment (XSEDE) supported by the National Science Foundation, under Grant No. ACI-1053575. The authors would also like to acknowledge extremely helpful discussions with Dr Sanjeev Kolli and Prof Anton van der Ven of the University of California Santa Barbara on the use of the CASM code.

\end{acknowledgement}


\bibliography{refs.bib}


\end{document}


\clearpage

\begin{figure}[htp!]
\centering
\includegraphics[width=1.0\textwidth]{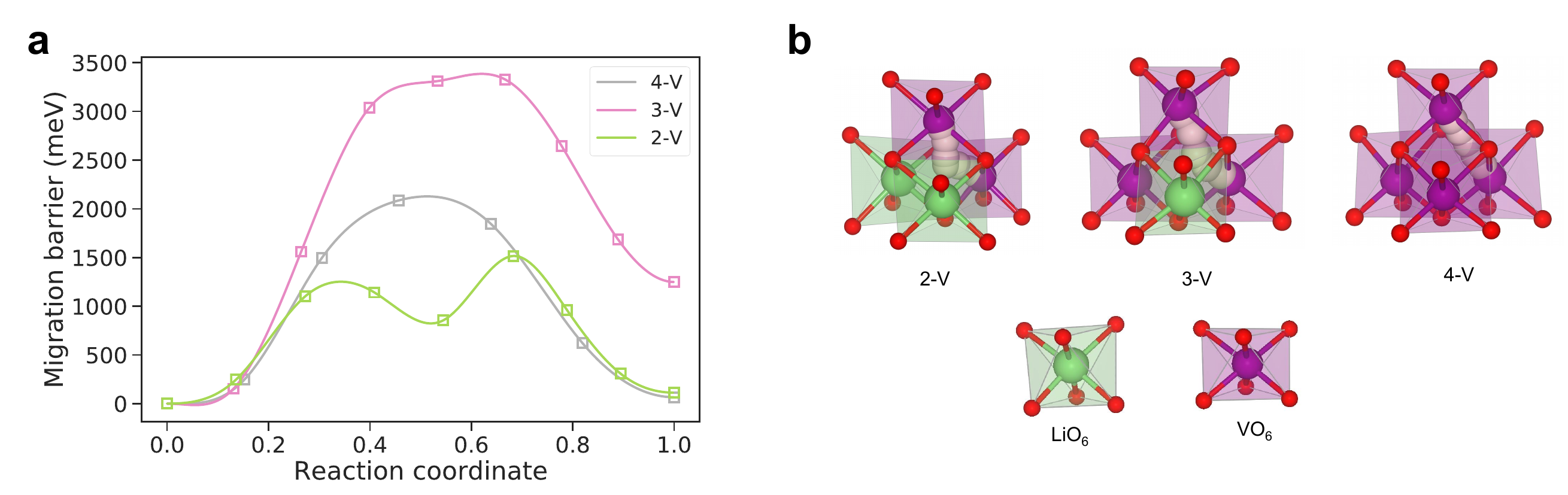}
\caption{\label{fig:neb_V} (a) NEB barriers for \ce{V^{4+}}/\ce{V^{3+}} migration through 2-V, 3-V and 4-V sites, (b) Illustration of the minimum energy path of V atoms in various local environments.}
\end{figure}

\begin{figure}[htp!]
\centering
{\includegraphics[width=1.0\textwidth]{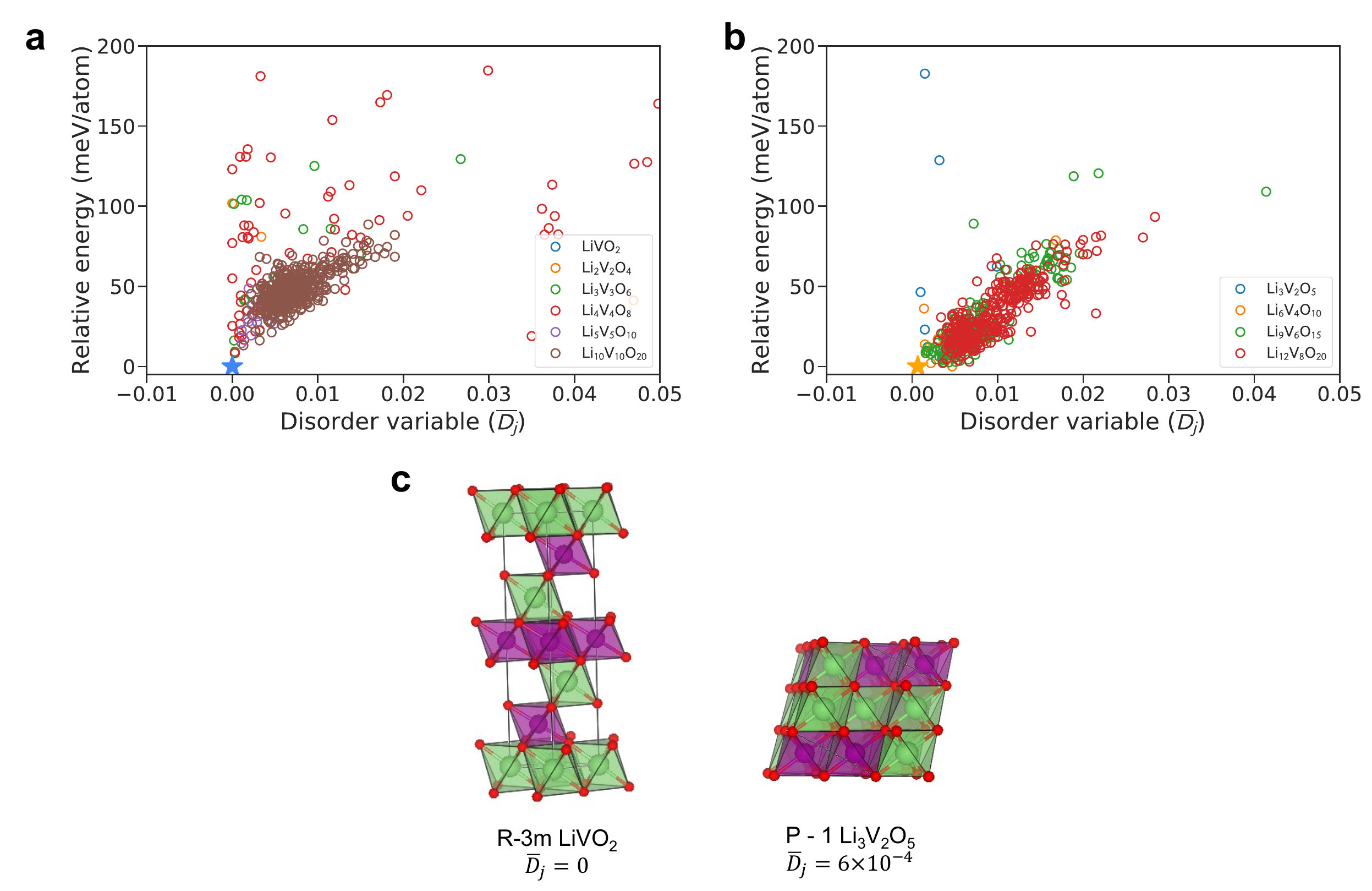}}
\caption{\label{fig:ss_cubic} (a) Relative energies of (a) \ce{LiVO2} and (b) \ce{Li3V2O5} structures as a function of order parameter ($\overline{D_j}$). The blue star in (a) represents $R\bar{3}m$ \ce{LiVO2} and the yellow star in (b) represents the \ce{Li3V2O5} structure with the lowest calculated energy. (c) Crystal structure of the calculated ground state $R\bar{3}m$ \ce{LiVO2} and \ce{Li3V2O5}.}
\end{figure}

\begin{figure}[htp!]
\centering
{\includegraphics[width=0.75\textwidth]{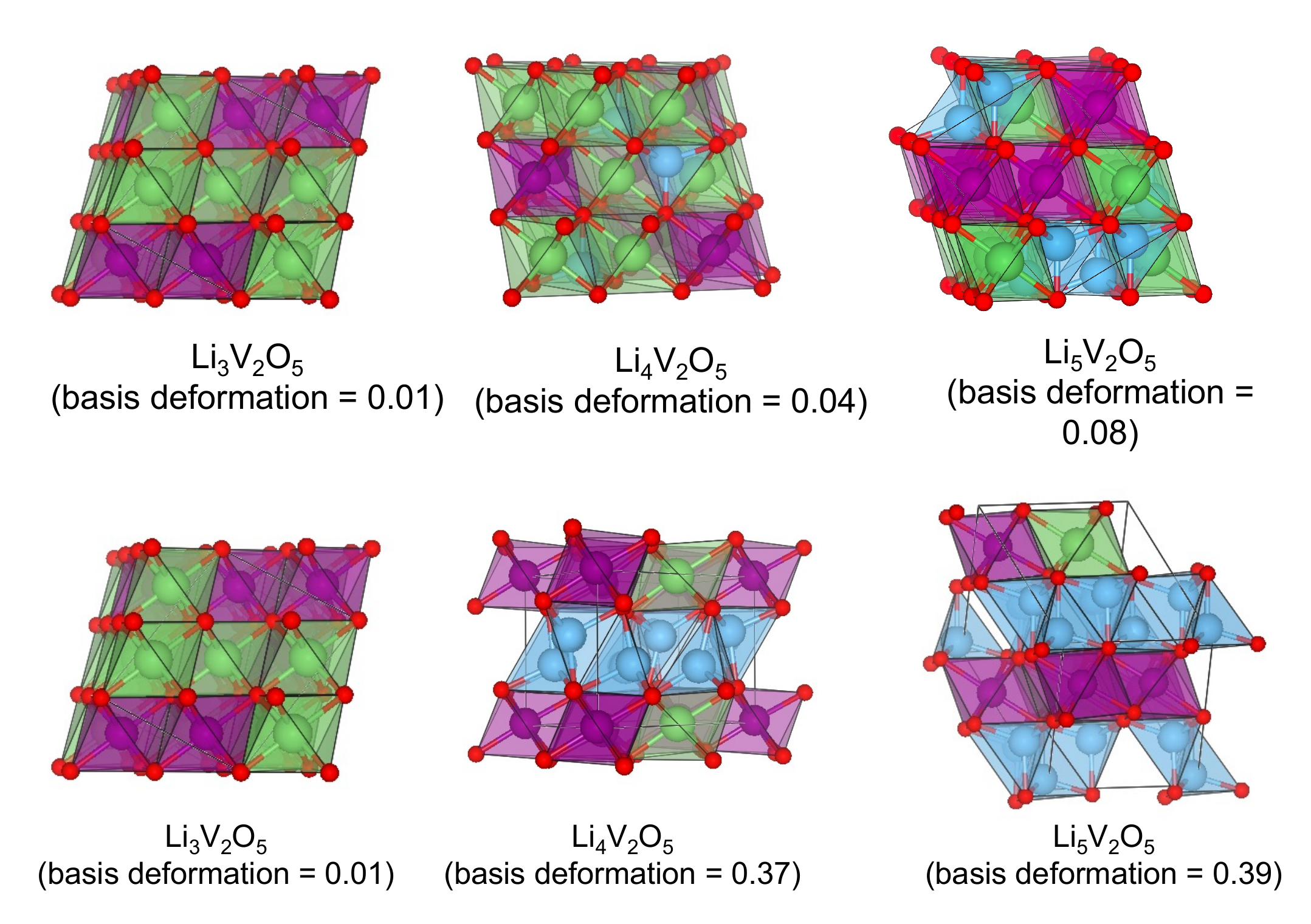}}
\caption{\label{fig:ss_cubic_1} The structures of lowest energy metastable cubic phase and the ground state of \ce{Li_{3+x}V2O5} (x = 0, 1, 2).}
\end{figure}

\begin{figure}[htp!]
\centering
\includegraphics[width=0.5\textwidth]{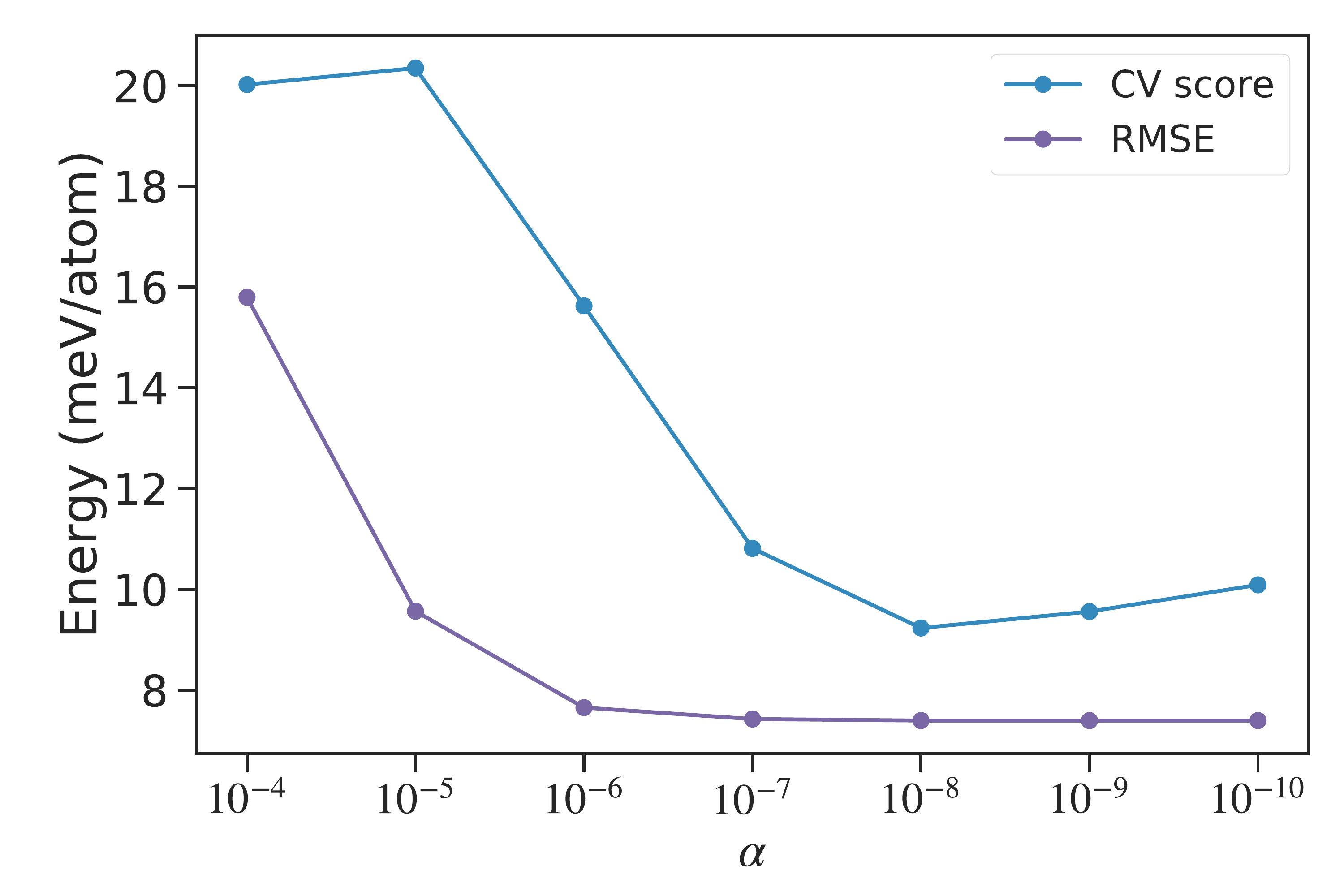}
\caption{\label{fig:alpha} Cross-validation (CV) score and root mean square error (RMSE) of the fitted cluster expansion model as a function of the regularization parameter ($\alpha$).}
\end{figure}

\clearpage

\begin{figure}[htp!]
\centering
{\includegraphics[width=0.6\textwidth]{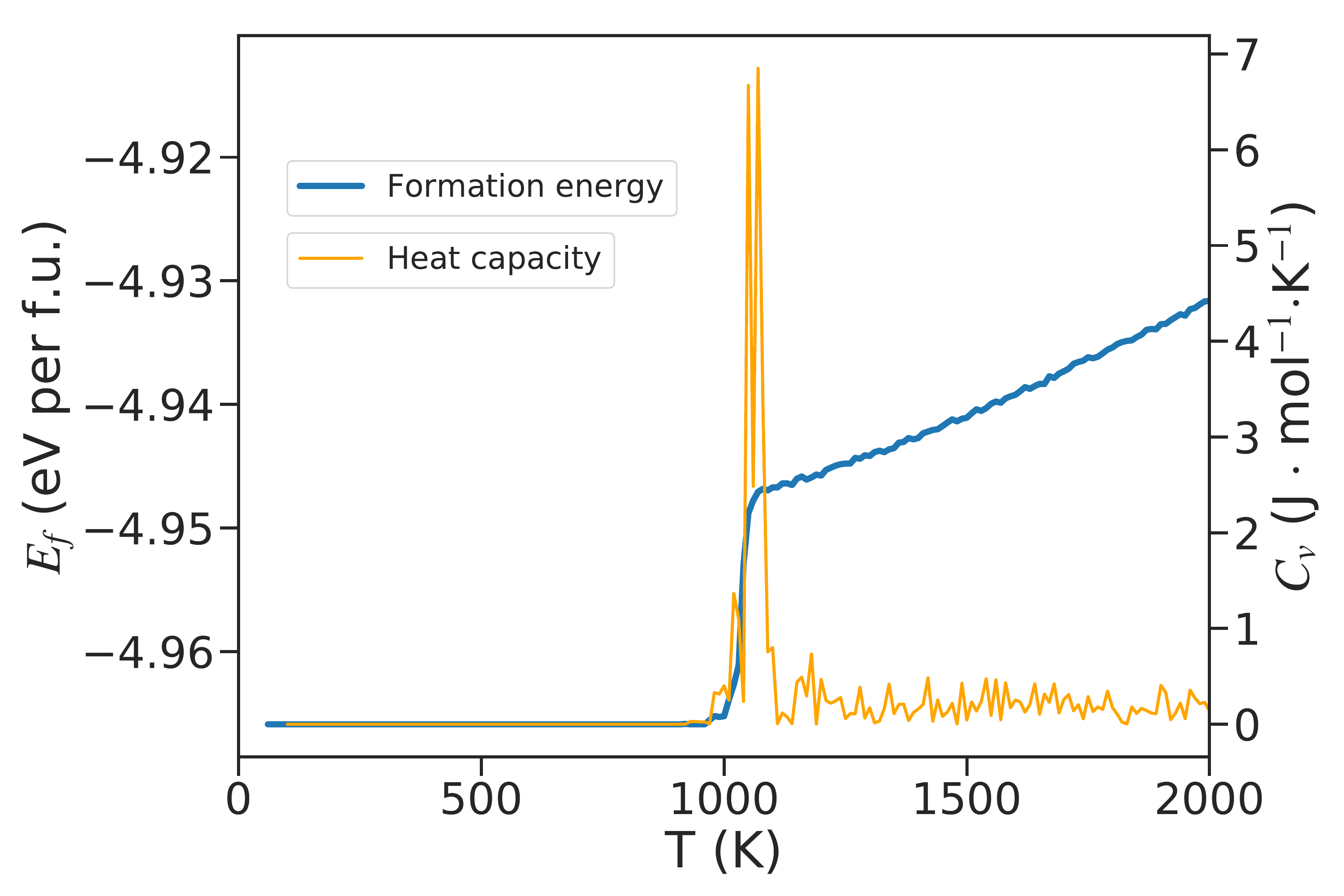}}
\caption{\label{fig:mc} Monte Carlo simulations of order-disorder phase transition of \ce{Li3V2O5} in the 10$\times$10$\times$10 supercell.}
\end{figure}

\begin{figure}[htp!]
\centering
\includegraphics[width=1.0\textwidth]{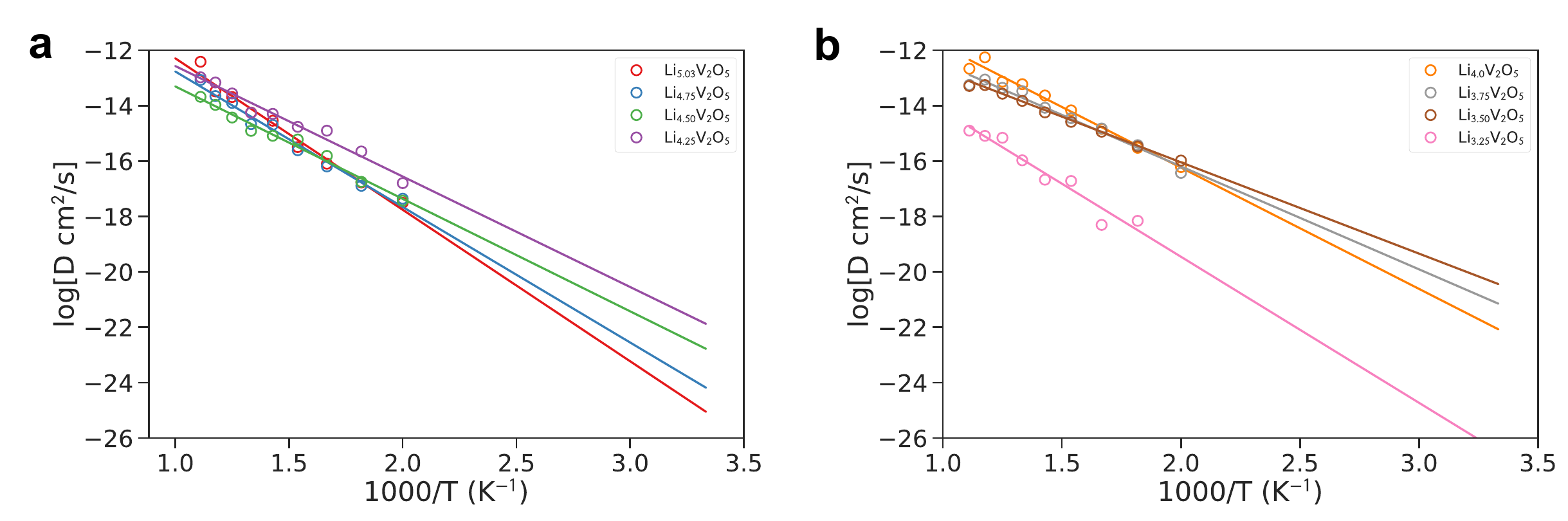}
\caption{\label{fig:arrhenius} Arrhenius plot from $NpT$ MD simulations of different  \ce{Li_{3+x}V2O5} compositions.}
\end{figure}

\clearpage

\begin{figure}[htp!]
\centering
\includegraphics[width=0.8\textwidth]{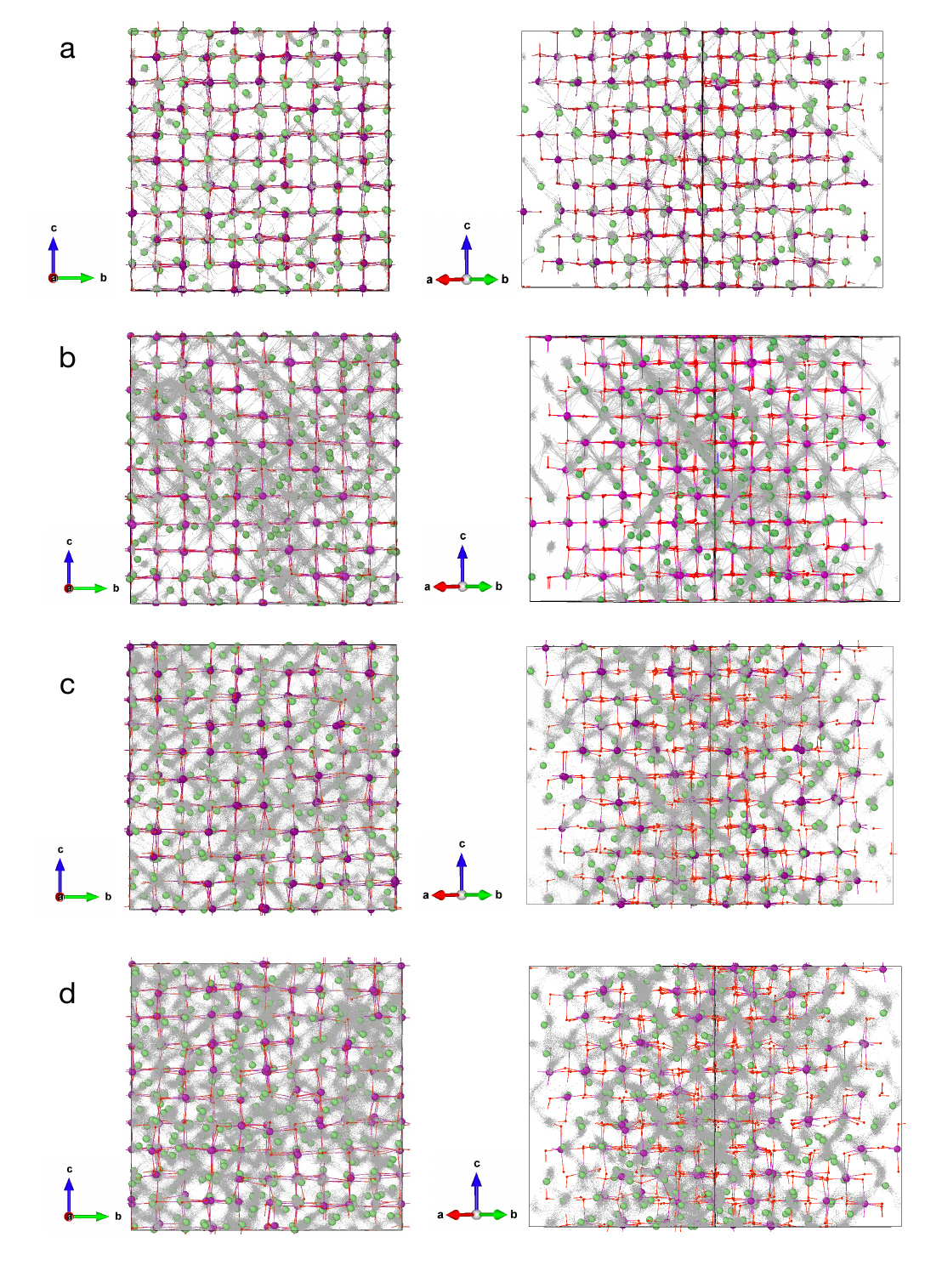}
\end{figure}
\clearpage
\begin{figure}[htp!]
\includegraphics[width=0.8\textwidth]{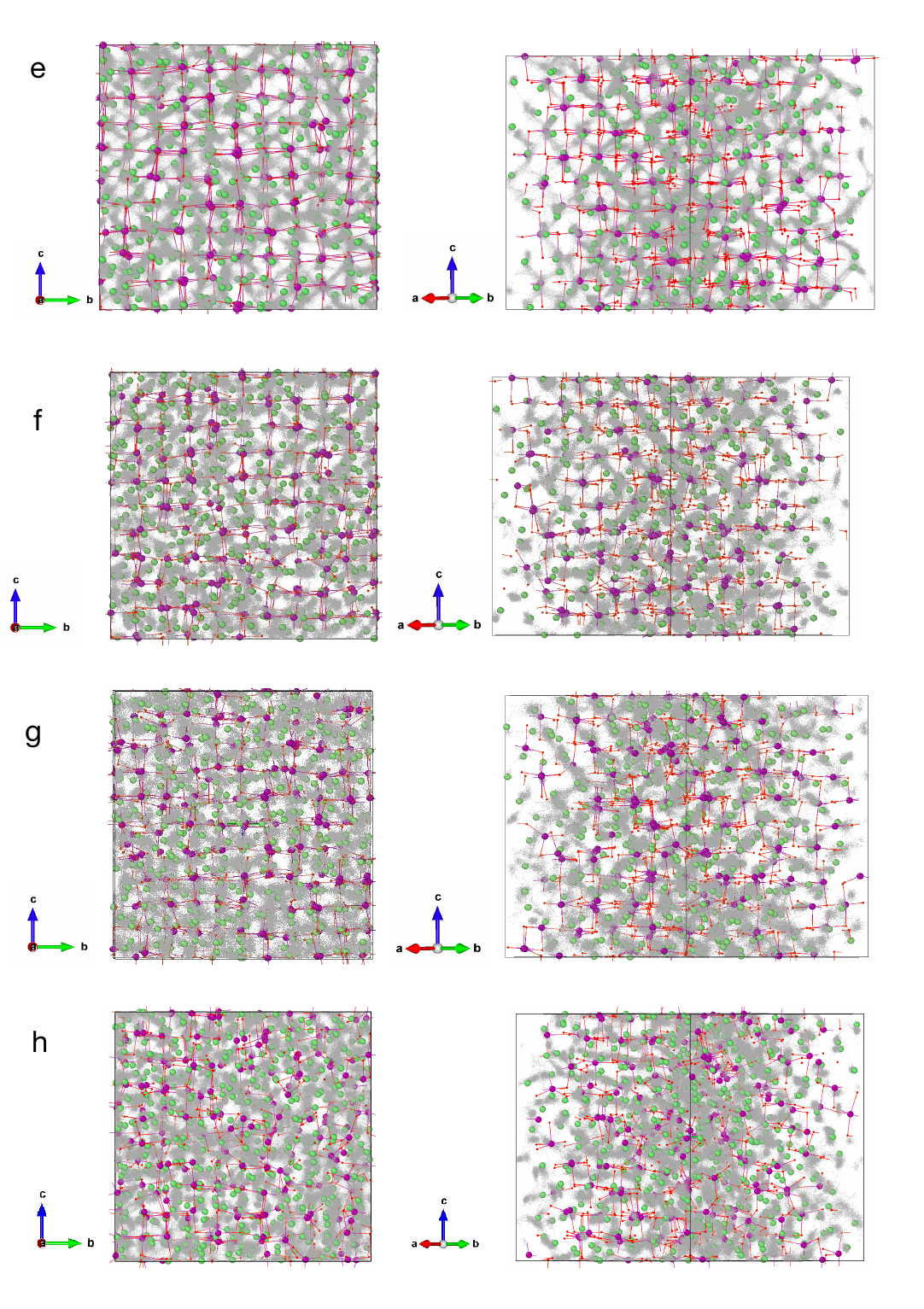}
\caption{\label{fig:trajectory} Calculated \ce{Li+} trajectories obtained from MD simulations. (a) \ce{Li_{3.25}V2O5}, (b) \ce{Li_{3.5}V2O5}, (c) \ce{Li_{3.75}V2O5}, (d) \ce{Li_{4.0}V2O5}, (e) \ce{Li_{4.25}V2O5}, (f) \ce{Li_{4.50}V2O5}, (g) \ce{Li_{4.75}V2O5} and (h) \ce{Li_{5.0}V2O5}.}
\end{figure}

\clearpage

\begin{figure}[htp!]
\centering
\includegraphics[width=1.0\textwidth]{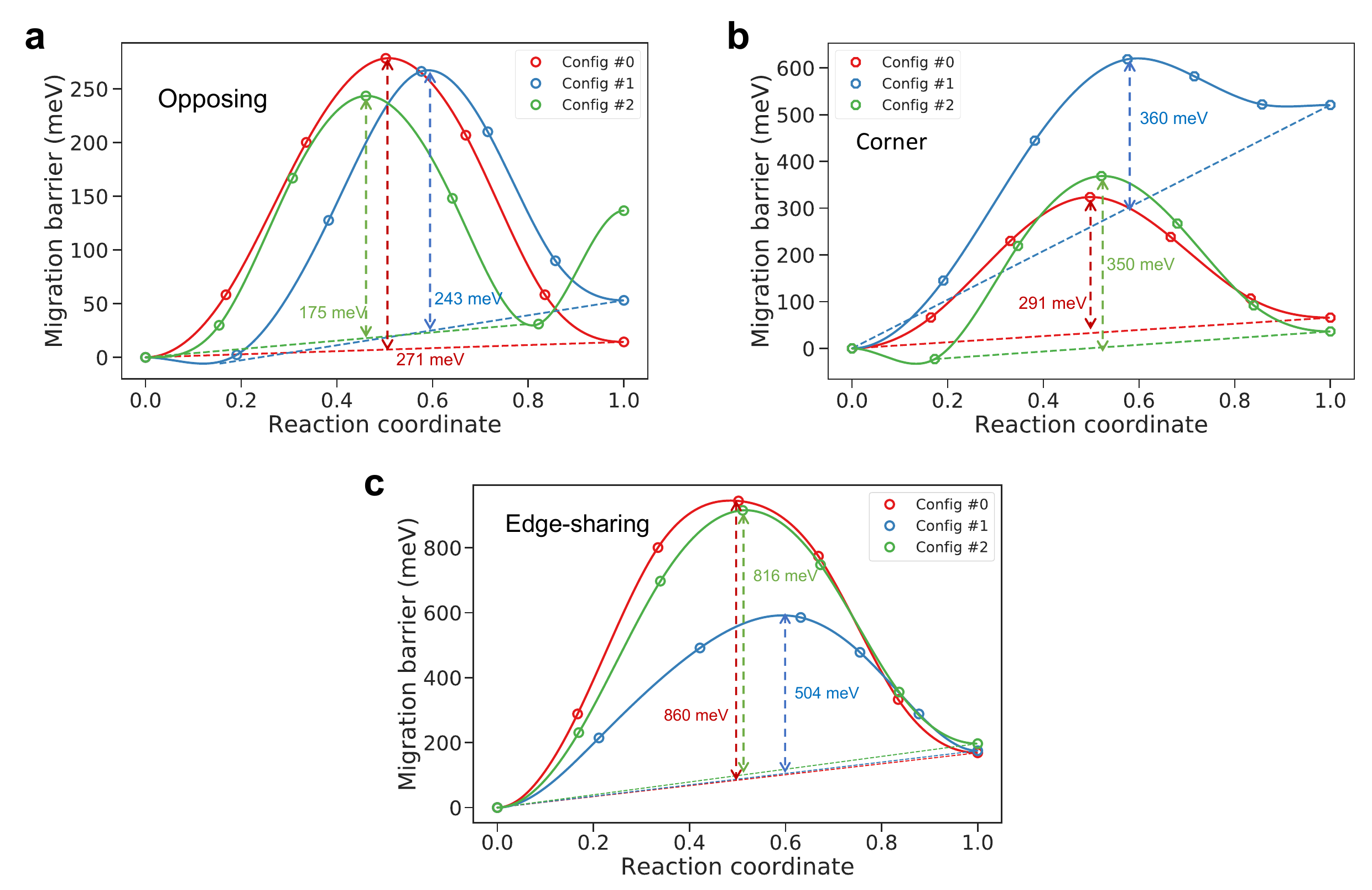}
\caption{\label{fig:neb_Li_0} DFT calculated \ce{Li+} migration barriers in 0-V sites. (a) 0-V to 0-V (Opposing), (b) 0-V to 0-V (Corner-Sharing) and (c) 0-V to 0-V (Edge-Sharing).}
\end{figure}
\clearpage

\begin{figure}[htp!]
\centering
\includegraphics[width=1.0\textwidth]{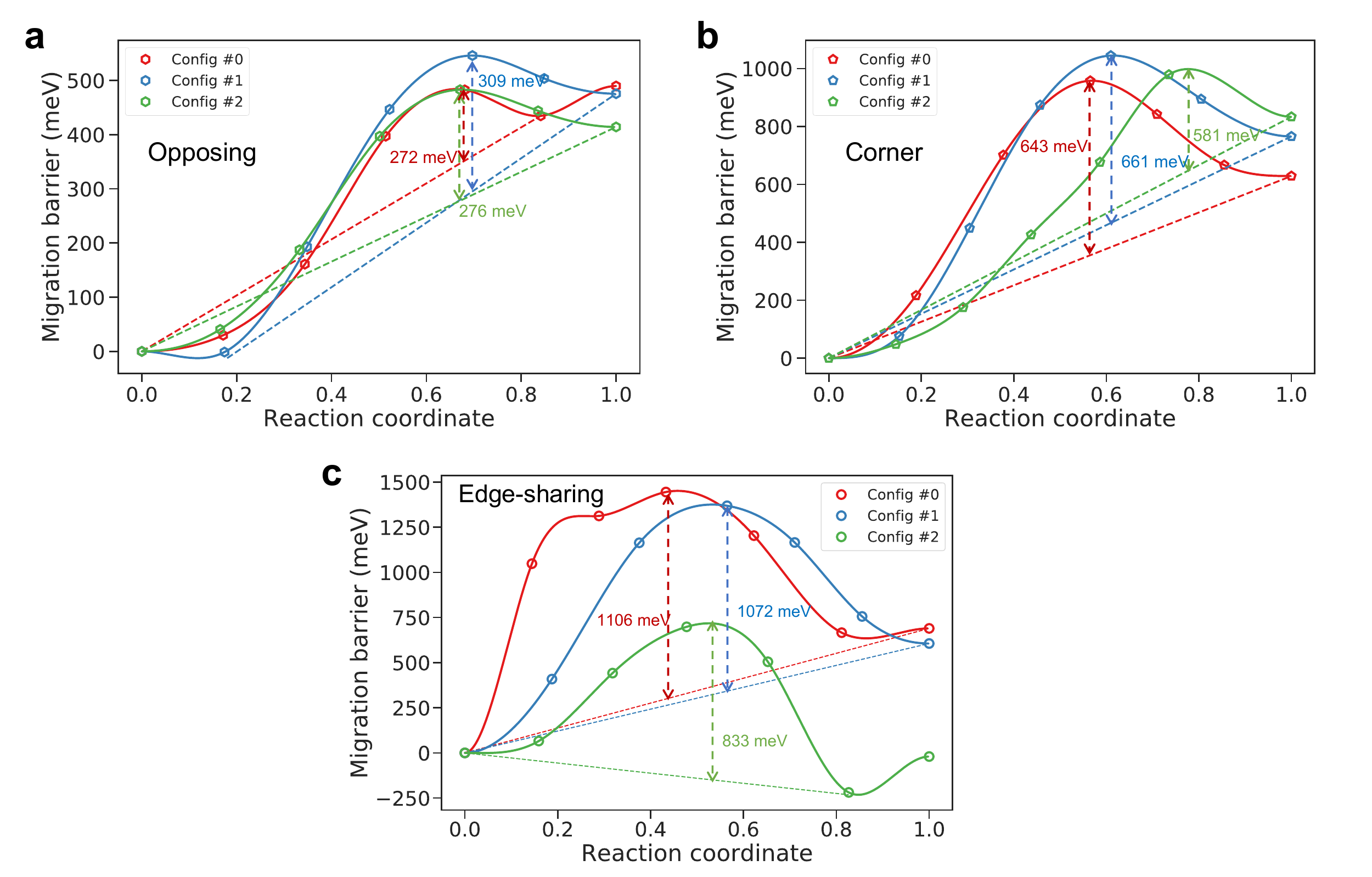}
\caption{\label{fig:neb_Li_0_1} DFT calculated \ce{Li+} migration barriers via 0-V and 1-V sites. (a) 0-V to 1-V (Opposing), (b) 0-V to 1-V (Corner-sharing) and (c) 0-V to 1-V (Edge-sharing).}
\end{figure}
\clearpage

\begin{figure}[htp!]
\centering
\includegraphics[width=1.0\textwidth]{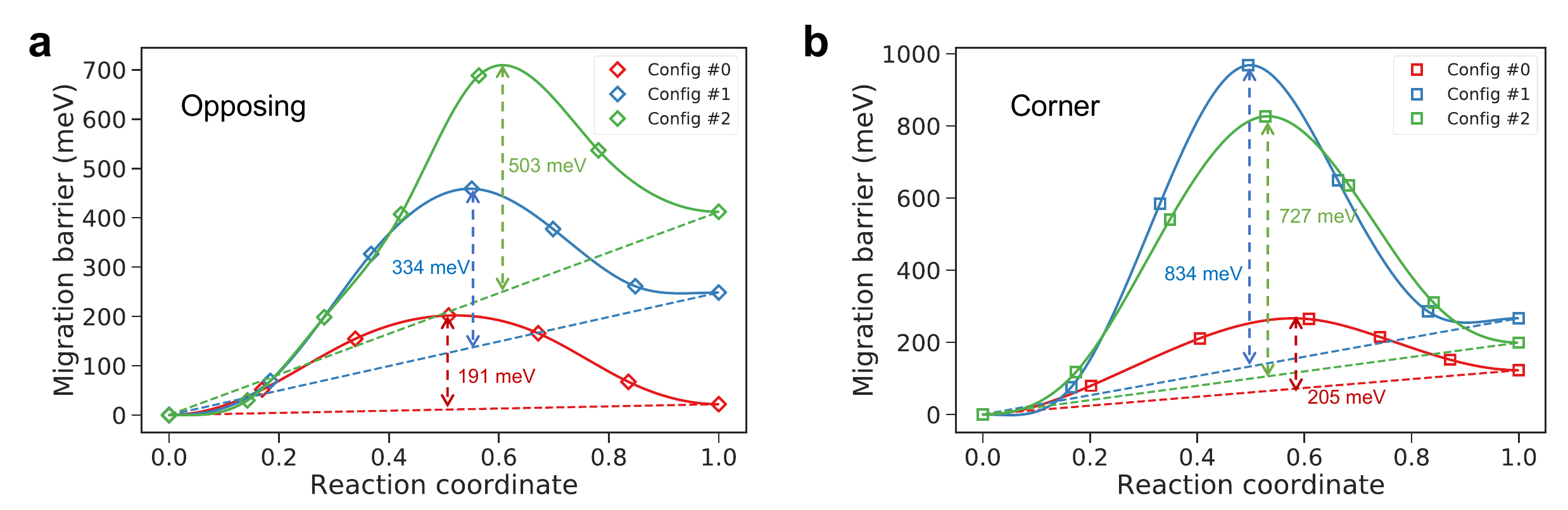}
\caption{\label{fig:neb_Li_1} DFT calculated \ce{Li+} migration barriers via 1-V sites. (a) 1-V to 1-V (Opposing), (b) 1-V to 1-V (Corner Sharing).}
\end{figure}

\begin{figure}[htp!]
\centering
\includegraphics[width=0.47\textwidth]{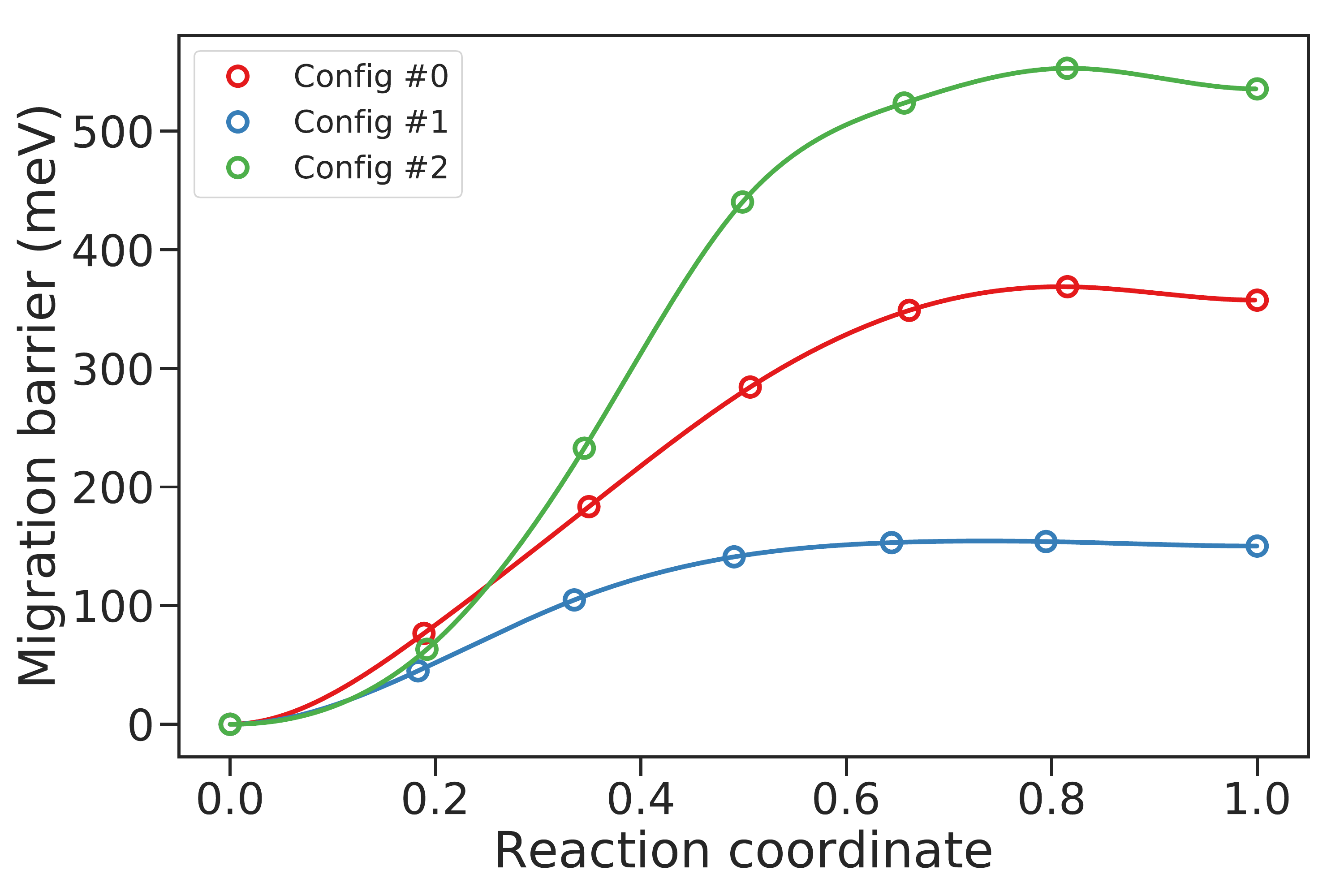}
\caption{\label{fig:neb_Li_2} DFT calculated \ce{Li+} migration barriers via connected 1-V and 2-V sites (Opposing).}
\end{figure}
\clearpage

\begin{figure}
    \centering
    \includegraphics[width=0.49\textwidth]{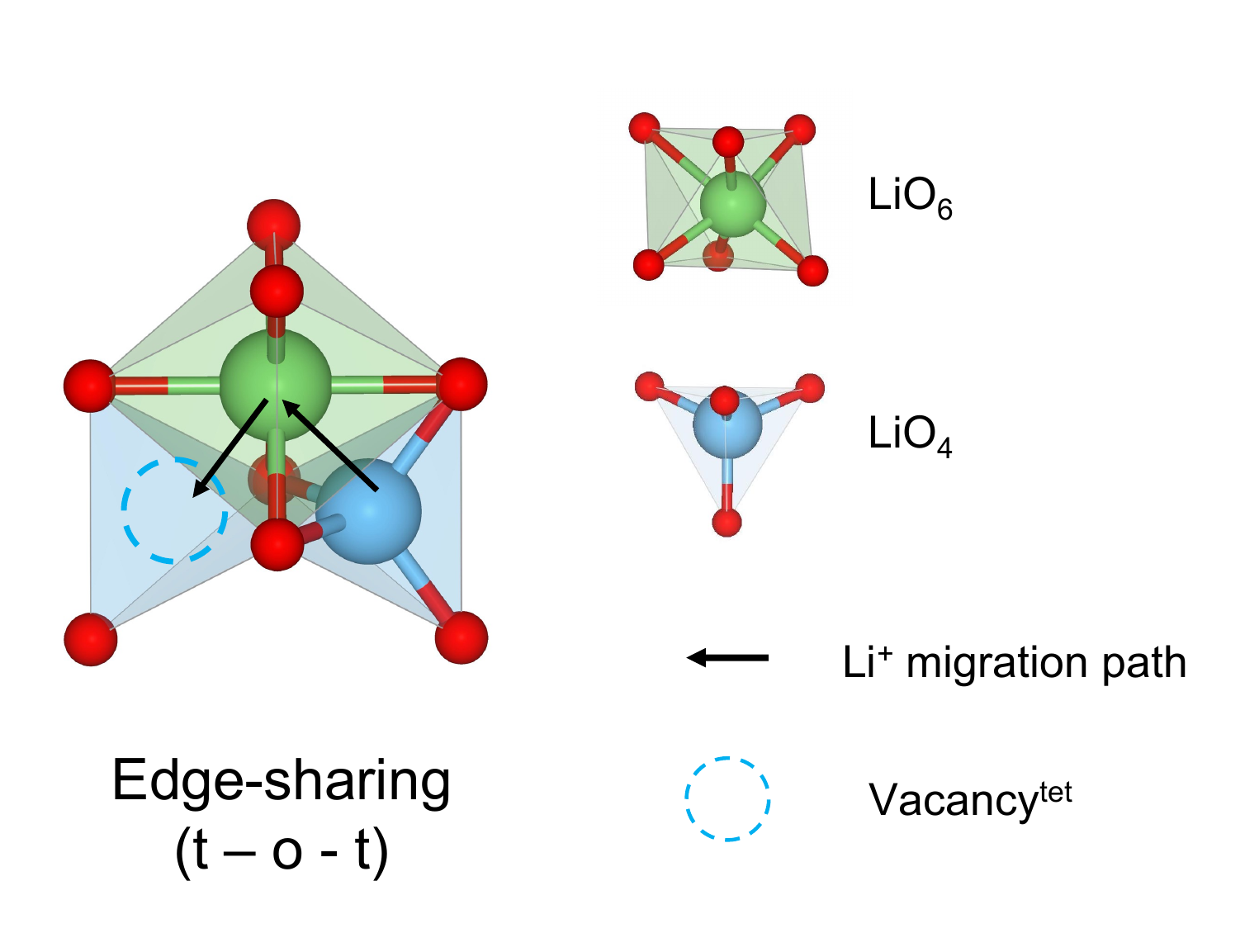}
    \caption{\label{fig:neb_edge_sharing} Illustration of \ce{Li+} migration via edge-sharing ``t-o-t " path in DRX-\ce{Li3V2O5}. }
\end{figure}

\begin{figure}
    \centering
    \includegraphics[width=0.49\textwidth]{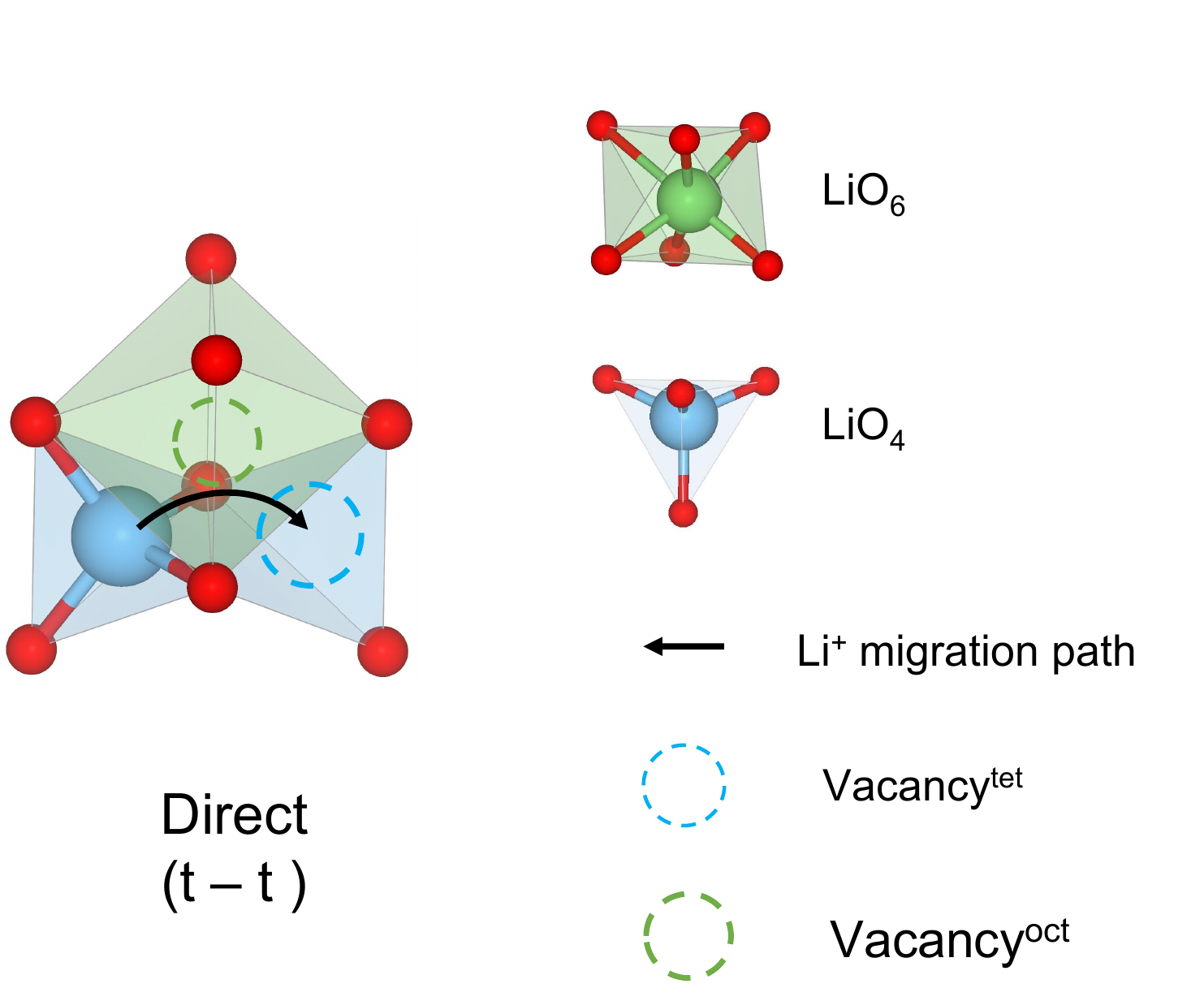}
    \caption{\label{fig:neb_t-t} Illustration of \ce{Li+} migration via ``t-t " path in DRX-\ce{Li3V2O5}. The direct pathway indicates a divacancy mechanism of \ce{Li^{tet}} vacancies at the end of discharge.}
\end{figure}